\documentclass{sig-alternate}
\usepackage{mathptmx} % This is Times font

\usepackage{fancyhdr}
\usepackage[normalem]{ulem}
\usepackage[hyphens]{url}
\usepackage[sort,nocompress]{cite}
\usepackage[final]{microtype}
\usepackage[keeplastbox]{flushend}

\usepackage{cite}
\usepackage{amsmath,amssymb,amsfonts}
\usepackage{algorithmic}
\usepackage{graphicx}
\usepackage{textcomp}
\usepackage{xcolor}
\usepackage{amsmath}
\usepackage{siunitx}
\usepackage[hyphens]{url}
\usepackage{booktabs}
\usepackage{subcaption}
\usepackage{multirow}

\usepackage{pifont}

\usepackage[bookmarks=true,breaklinks=true,letterpaper=true,colorlinks,citecolor=blue,linkcolor=blue,urlcolor=blue]{hyperref}

\newcommand{\M}{SupeRBNN}
\newcommand{\rev}[1]{#1}
\newcommand{\revm}[1]{#1}
\title{\M: Randomized Binary Neural Network Using Adiabatic Superconductor Josephson Devices}
\author{Zhengang Li\textsuperscript{1}, Geng Yuan\textsuperscript{2}, Tomoharu Yamauchi\textsuperscript{3}, Zabihi Masoud\textsuperscript{1}, Yanyue Xie\textsuperscript{1}, Peiyan Dong\textsuperscript{1},\\
Xulong Tang\textsuperscript{4}, Nobuyuki Yoshikawa\textsuperscript{5}, Devesh Tiwari\textsuperscript{1}, Yanzhi Wang\textsuperscript{1}, Olivia Chen\textsuperscript{3} \\
\textsuperscript{1}Northeastern University, 
\textsuperscript{2}University of Georgia, 
\textsuperscript{3}Tokyo City University, \\
\textsuperscript{4}University of Pittsburgh, 
\textsuperscript{5}Yokohama National University \\
{E-mail:} \textsuperscript{1}\{li.zhen, m.zabihi, xie.yany, dong.pe, d.tiwari, yanz.wang\}@northeastern.edu, \\
\textsuperscript{2}geng.yuan@uga.edu,
\textsuperscript{3}g2281461@tcu.ac.jp, olivia.chen@ieee.org, \\
\textsuperscript{4}tax6@pitt.edu,
\textsuperscript{5}nyoshi@ynu.ac.jp} 

\begin{document}
\maketitle
\pagestyle{plain}

\newcommand{\GY}[1]{\textcolor{blue}{Geng: #1}}
\newcommand{\todo}[1]{\textcolor{red}{\sf\bfseries Todo: #1}}
\newcommand{\bred}[1]{\textcolor{red}{\sf\bfseries #1}}
\newcommand{\red}[1]{\textcolor{red}{#1}}
\newcommand{\blue}[1]{\textcolor{blue}{#1}}
\newcommand{\yellow}[1]{\textcolor{yellow}{#1}}
\newcommand{\purple}[1]{\textcolor{purple}{#1}}
\newcommand{\brown}[1]{#1}
\newcommand{\cross}[1]{\textcolor{red}{\sout{#1}}}

%%%%%% -- PAPER CONTENT STARTS-- %%%%%%%%

\begin{abstract}
Adiabatic Quantum-Flux-Parametron (AQFP) is a superconducting logic with extremely high energy efficiency. By employing the distinct polarity of current to denote logic `0' and `1', AQFP devices serve as excellent carriers for binary neural network (BNN) computations.  
% The principle of AQFP buffer using different directions of current to present logic '0' and '1' makes AQFP devices good carriers for binary neural network (BNN) computation. 
% The Binarized Neural Network (BNN),
% with minimal memory requirements and no reliance on multiplication
% The principle of the AQFP devices makes them suitable for crossbar-based in-memory computing. 
\rev{Although recent research has made initial strides toward developing an AQFP-based BNN accelerator, several critical challenges remain, preventing the design from being a comprehensive solution. In this paper, we propose~\M, an AQFP-based randomized BNN acceleration framework that leverages software-hardware co-optimization to eventually make the AQFP devices a feasible solution for BNN acceleration. Specifically, we investigate the randomized behavior of the AQFP devices and analyze the impact of crossbar size on current attenuation, subsequently formulating the current amplitude into the values suitable for use in BNN computation. To tackle the accumulation problem and improve overall hardware performance, we propose a stochastic computing-based accumulation module and a clocking scheme adjustment-based circuit optimization method.}
% The stochastic computing-based accumulation module and the clocking scheme adjustment-based circuit optimization are proposed to handle the accumulation problem and improve the overall hardware performance, respectively. 
% Intriguingly, we discover that the randomized behavior observed in the AQFP buffer can be seamlessly translated to the stochastic computing domain via a specific observation window with minimal hardware overhead. 
\rev{To effectively train the BNN models that are compatible with the distinctive characteristics of AQFP devices, we further propose a novel randomized BNN training solution that utilizes algorithm-hardware co-optimization, enabling simultaneous optimization of hardware configurations. In addition, we propose implementing batch normalization matching and the weight rectified clamp method to further improve the overall performance.
We validate our~\M~framework across various datasets and network architectures, comparing it with implementations based on different technologies, including CMOS, ReRAM, and superconducting RSFQ/ERSFQ. Experimental results demonstrate that our design achieves an energy efficiency of approximately $7.8 \times 10^4$ times higher than that of the ReRAM-based BNN framework while maintaining a similar level of model accuracy. Furthermore, when compared with superconductor-based counterparts, our framework demonstrates at least two orders of magnitude higher energy efficiency.}

\end{abstract}

\section{Introduction} \label{sec:intro}
In recent years, deep learning and deep neural networks (DNNs) have become the core enabler of a broad spectrum of artificial intelligence (AI) applications such as image recognition~\cite{deng2009imagenet}, natural language processing~\cite{devlin2019bert}, and autonomous driving~\cite{bechtel2018deeppicar}. 
However, the high computation and storage demands of DNN executions are still an essential challenge for the democratization of AI.

A significant amount of effort has been dedicated to reducing network energy consumption at the algorithmic level. Recent studies have proposed Binary Neural Networks (BNNs) as a solution~\cite{courbariaux2015binaryconnect,courbariaux2016binarized,rastegari2016xnor,lin2017towards}, which have a 32$\times$ smaller memory footprint than conventional DNNs that use 32-bit floating-point precision, despite having a similar network structure.
% For example, all the weights in BNNs are set to either +1 or -1.
% In this way, with a similar network structure, BNNs inherently require 32$\times$ less memory footprint than the conventional DNNs that use the 32-bit floating-point precision.
Additionally, BNNs can avoid the tremendous floating-point multiply-accumulation (MAC) operations in conventional DNN models by employing bit-wise exclusive-NOR and popcount logic.
% Moreover, BNNs can also avoid the tremendous floating-point multiply-accumulation (MAC) operations in DNN models by simply using bit-wise exclusive-NOR and popcount logics.
% thanks to the 1-bit representation.
% Moreover, BNNs can also avoid the tremendous floating-point multiply-accumulation (MAC) operations introduced by the matrix multiplications, which dominate the computations of neural networks. 
% Instead, the BNNs can implement those floating-point MAC operations by simply using bit-wise exclusive-NOR and popcount logics, thanks to the 1-bit representation.
\brown{In recent years, advancements in BNN training techniques and network structure designs have led to significant improvements in network accuracy~\cite{xu2021recu,bulat2020bats}, making BNN a promising candidate for energy-efficient-oriented designs.}

In addition to algorithmic optimizations, significant advancements have been achieved in the hardware domain, with superconducting electronics (SCE) being a prime example.
% \rev{Owing to the high speed and low energy dissipation of superconducting logic devices, several studies investigate efficient neural network accelerator implementations based on superconducting technologies, such as superNPU~\cite{ishida2020supernpu} and JBNN~\cite{fu2022jbnn}.
Superconducting logic families, leveraging magnetic flux quantization and quantum interference in Josephson-junction (JJ)-based superconductor loops, have emerged as promising candidates for future computing. 
% Superconducting logic families utilizing magnetic flux quantization and quantum interference in Josephson-junction (JJ)-based superconductor loops have been considered promising candidates for future computing technologies.
The IEEE International Roadmap on Devices and Systems (IRDS) has recognized SCE as one of the top-level roadmaps since 2018~\cite{debenedictis2018ieee,holmes2021cryogenic}.
Among various superconducting logic families, Adiabatic Quantum-Flux-Parametron (AQFP) logic stands out for its exceptional energy efficiency. 
In 2019, researchers experimentally demonstrated a 1.4 zJ energy dissipation for each operation in AQFP at the device level~\cite{1.4zJ}. On the circuit level, authors in~\cite{chen2019adiabatic} have shown that, compared to state-of-the-art CMOS technology, AQFP can achieve an energy-efficiency gain in the range of $10^4\sim10^5$.In addition, research on AQFP design automation has been conducted worldwide, aiming to achieve system-level AQFP circuit design and implementation \cite{cai_glsvlsi,cai_iccd,epfl_aspdac,saito_tas2021}. Thanks to advancements in the EDA environment for AQFP VLSI design, several successful AQFP chips have been demonstrated \cite{tas2019,MANA,Yamae_2019}.

\brown{
Diverging from previous neural network acceleration efforts focused on RSFQ superconducting logic, such as superNPU~\cite{ishida2020supernpu} and JBNN~\cite{fu2022jbnn}, recent research~\cite{10040729} has recognized the immense potential of integrating BNN with AQFP technology to achieve exceptionally efficient DNN accelerator design, marking an initial endeavor in this direction. In~\cite{10040729}, a crossbar synapse array architecture using AQFP devices is proposed, which is a prototype module that intends to efficiently compute the vector-matrix multiplications required for the MAC operation in BNNs.
However, this is far from a complete solution to make the AQFP devices feasibly used for BNN acceleration. There are still several critical challenges that need to be addressed.
% Recent work~\cite{10040729} proposes a crossbar synapse array architecture using AQFP devices, as a prototype module, that can be used to compute the vector-matrix multiplications of MAC operation in BNNs.
}

\brown{First of all, the utilization of AQFP devices for constructing crossbar arrays poses a challenge regarding their \emph{randomized behavior issue} (\textbf{Challenge \#1}).
Specifically, when building an AQFP-based crossbar, the accumulated current in the analog domain on each crossbar column will suffer from the \emph{current attenuation} caused by the increasing superconductive inductance as the crossbar size increases.
With an attenuated input current, the AQFP buffer may not be able to precisely detect the direction of the input current, resulting in randomized outputs (more details in Section~\ref{sec:current_attenuation_analysis}).
Such randomness will introduce inaccuracy in BNN computation, 
resulting in significant discrepancies between the BNN model that is trained on software and its actual behavior during hardware implementation.
Consequently, the accuracy of the network may degrade substantially.}
% As a result, the network's accuracy can degrade significantly.

\brown{Moreover, due to the current attenuation issue and immature manufacturing technology, the AQFP-based crossbar has limited scalability (\textbf{Challenge \#2}). This indicates that the size of the crossbar array cannot be arbitrarily large. Multiple crossbar arrays must be employed to accommodate all the weights of a BNN layer or a convolutional filter.
However, this will raise another problem: how to effectively accumulate the intermediate results from the corresponding crossbar columns between multiple crossbars (\textbf{Challenge \#3}).
This is not a trivial task since the binary intermediate results are used. Inappropriately addressing this problem can lead to significant accuracy degradation.
Last but not least, the hardware configurations such as crossbar size and the threshold current of the AQFP buffer also need to be optimized to deliver the best accuracy while considering the hardware performance (\textbf{Challenge \#4}).}

% Is this yet another crippled design that has to end with compromised accuracy or hardware performance?
\brown{Due to these critical challenges, we would like to ask whether this is yet another crippled design that has to end with compromised accuracy or hardware performance?} 
\rev{Fortunately, the answer is {\em no}. To overcome these four challenges, we first investigate the randomized behavior of AQFP devices. Then, we propose an AQFP randomized behavior-aware BNN training paradigm, which incorporates the randomized behavior of AQFP buffer by formulating the binarization process of the output feature maps in a probabilistic manner according to the amplitude of the crossbar's output current (\textbf{Contribution \#1}). We also incorporate the weight-rectified clamp method to help improve randomized BNN training accuracy.
After that, we simulate the impact of current attenuation in terms of different crossbar sizes and formulate the current amplitude into the mathematical value used in BNN computation. 
By doing this, the gap between the BNN model that is trained on software and the model's actual behavior in hardware implementation can be mitigated (\textbf{Contribution \#2}). }
% By doing this, we can directly incorporate the current amplitude within crossbars as the latent activation in our training process and mitigate the gap between software trained model with hardware implementation

\rev{
Intriguingly, we find that the unique randomness behavior of AQFP devices is inherently compatible with the stochastic computing (SC) technique.
Therefore, to solve the problem of accumulating the intermediate results from the corresponding crossbar columns between multiple crossbars, we propose a novel and efficient SC-based accumulation module circuit to add up the intermediate result as well as improve the model accuracy impacted by the randomized behavior (\textbf{Contribution \#3}).
Since the randomized behavior that appears in the AQPF buffer's output is constrained and dependent on the input current amplitude, it can be seamlessly converted to a stochastic number (SN) via a specific observation window with minimal hardware overhead.
% Since the unique randomness characteristic makes the AQFP inherently compatible with the stochastic computing (SC) technique, to effectively accumulate the intermediate results from the corresponding crossbar columns between multiple crossbars, we propose a novel and efficient SC-based accumulation module circuit to add up the intermediate result as well as improve the model accuracy impacted by the randomized behavior (\textbf{Contribution \#3}). 
Due to the significant influence of hardware configurations on the model accuracy, we propose a comprehensive software-hardware co-optimization. 
This helps optimize the hardware configurations of AQFP-based BNN accelerator design, such as crossbar synapse array size, stochastic computing bit-stream length, and ``gray-zone'' width of AQFP buffer by comprehensively considering power consumption, energy efficiency, and hardware computing error (\textbf{Contribution \#4}).}
\rev{Besides that, we introduce a batch normalization (BN) matching method to address the floating-point computation problem induced by BN layer with no additional peripheral circuits. And a clocking scheme adjustment-based circuit optimization is also proposed to improve the hardware performance (\textbf{Contribution \#5}).}
% \rev{\textit{In summary, we investigate and simulate the randomized behavior and current attenuation feature of the AQFP-based crossbar architecture, and propose the randomized-aware BNN training algorithm as well as the algorithm-hardware co-optimization effectively integrating the randomized behavior into the BNN training process, achieving the best-suited hardware configuration and closing the gap between the software-trained model with the actual hardware implementation.}}

\rev{To validate the effectiveness of~\M, a series of detailed comparative experiments are provided. We analyze the accuracy distribution according to multiple hardware configurations and the sensitivity of the relationship between SC bit-stream length with model accuracy. We also compare our method with multiple representative technologies, including CMOS, ReRAM, and RSFQ/ERSFQ on MNIST and CIFAR-10 datasets. 
~\M~achieves about $7.8\times10^4$ times higher energy efficiency with a similar model accuracy level compared with the representative ReRAM-based BNN framework on CIFAR-10 dataset.}

\section{Background and Related Work} \label{sec:related}

\subsection{Model Quantization and Binary Neural Network}

Model quantization is a crucial technique for DNN inference acceleration.
% which has been intensively explored for deep neural networks (DNNs) to compress the model size and improve the inference speed. 
It maps the 32-bit floating-point weight and activation values in a DNN model using fewer bits representation.
% to lower bit-width values, by splitting the data range into a certain number of levels. 
% ~\cite{gong2014compressing} proposed to quantize the weights of fully connected layers in a deep network by vector quantization techniques and proved that high-precision parameters are not necessary for achieving high performance in deep networks. 
Existing model quantization research can be categorized according to quantization schemes. 
Binary neural network (BNN)~\cite{courbariaux2015binaryconnect,courbariaux2016binarized,rastegari2016xnor,lin2017towards} and ternary neural network (TNN)~\cite{zhu2016trained,he2019simultaneously} use extremely low precision for DNN models, and low-bit-width fixed-point neural network~\cite{zhou2016dorefa, choi2018pact} quantizes models with the same interval between each quantization level. Among them, with weights constrained to $\{-1, 1\}$, multiplications of BNN can be replaced by additions/subtractions. Additions/subtractions can also be eliminated using XNOR and AND operations if activations are quantized to binary as well. This can significantly reduce operations and simplify hardware implementation, which is ideal for low-power consumption scenarios. 
% For example, based on reports from above works, accuracy typically degrades by $>5\%$ under binary scheme, and $2-3\%$ for ternary quantization.

As a pioneer work, Courbariaux et al.~\cite{courbariaux2016binarized} first binarized both weights and activations with the sign function. To overcome the almost everywhere zero gradients in the sign function, they incorporated the STE~\cite{bengio2013estimating} as an approximation to enable the gradient back-propagation. 
However, the limited representational ability of BNNs leads to a significant drop in accuracy. 
To mitigate the accuracy drop, XNOR-Net~\cite{rastegari2016xnor} introduces scaling factors obtained from the $L_1$-norm of the weights or activations to reduce the quantization error. 
% After that, XNOR-Net++~\cite{bulat2019xnor} fuses scaling factors of quantized weights and activations together as a learnable parameter and trains it via the standard back propagation. 
Then, the rotated binary neural network (RBNN)~\cite{lin2020rotated} explores and reduces the quantization error by considering the influence of the angular bias between the binarized weight vector and its full-precision version. 
Later works propose new gradient estimation functions and binarization-friendly network architectures to promote the BNN performance~\cite{gong2019differentiable,yang2019quantization,qin2020forward,lin2017towards,liu2020reactnet,leng2018extremely,xu2021recu}.

\subsection{AQFP Superconducting Logic}
% The standard cell library of AQFP includes basic logic gates such as BUFFER, INVERTER, AND, OR, MAJORITY, and SPLITTER.
% An AQFP logic gate is driven by AC-power, which serves both as the excitation current and synchronization mechanism.
AQFP originates from quantum-flux-parametron (QFP) logic, one among many superconducting logic families, which was first proposed in 1985 \cite{qfp}. Authors in \cite{aqfp} proposed an adiabatic version of QFP to obtain extremely low energy dissipation by re-parameterizing the device to allow QFP gates to be operated at an adiabatic mode, resulting in roughly 5$\sim$6 orders lower energy dissipation than its CMOS counterpart. 
% Furthermore, recent researches on the reversible version of QFP (RQFP) suggest that the energy required for bit-level information transfer can even go below the Shannon limit ($k_BTln2$) \cite{rqfp}. 

\begin{figure}[t]
  \centering
  \includegraphics[width=0.7\linewidth]{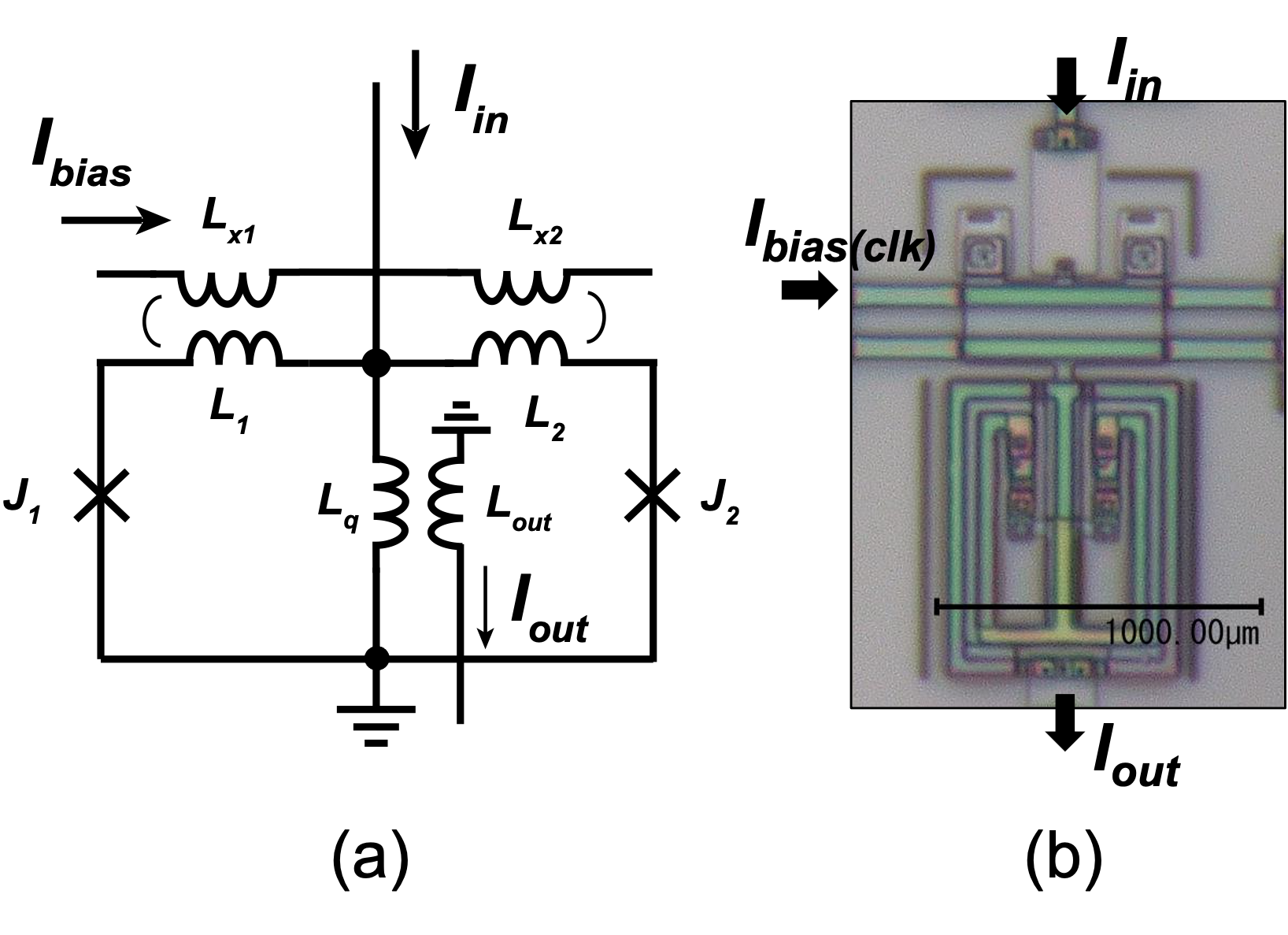}
  % \vspace{-1em}
  \caption{Adiabatic Quantum-Flux-Parametron Logic. (a) Schematic of an AQFP buffer. (b) Microphotograph of the fabricated AQFP buffer using 4-layer niobium process~\cite{hstp}. }
  \label{fig:AQFP_basic}
  % \vspace{-2em}
\end{figure}

Like many other superconductor-based logic families, AQFP also employs the Josephson Junction (JJ) as the basic switching element to obtain the state transition for logic encoding.
The most basic structure of AQFP circuits is the AQFP buffer, which consists of a double-Josephson-Junction SQUID ($J_1$, $J_2$)~\cite{clarke2006squid}, as shown in Figure~\ref{fig:AQFP_basic}.
A minimalist approach has been proposed to create an AQFP cell library containing essential logic gates (e.g., INVERTER, AND, OR, and MAJORITY gates) built from AQFP buffers for digital circuit design \cite{minimalist}.
Utilizing different directions (positive and negative) of output current pulses ($I_{out}$) to represent distinct logic states (0 or 1), the accumulation operation for outputs from various AQFP gates can be efficiently achieved through a straightforward current summation in the analog domain.
Moreover, when keeping a high excitation current to an AQFP buffer, the logic state stored in the AQFP buffer can be retained, making it possible to be used as a single-bit memory cell for storing the 1-bit BNN weights.
These characteristics render AQFP well-suited for addressing MAC operations in BNNs using a crossbar-based in-memory computing architecture.

However, due to the principle of AQFP buffer, the output current is sensitive to the direction of the input current. \rev{When the amplitude of input current is very small, which falls in the ``grayzone'' $\Delta I_{in}$ \cite{filippov1995} of an AQFP buffer,} the stochastic switching behavior (caused by the thermal or quantum fluctuation) exists in an AQFP buffer will make it hard to detect the direction of the input current, resulting in a randomized output with a probability related to input current, i.e., $0<P(I_{in})<1$.
This unique property is a double-edged sword: it introduces inaccuracy but also enables compatibility with stochastic computing.
% This unique property is a double-sided sword that introduces inaccuracy but also makes it possible to combine with stochastic computing.

\rev{Diverging from the previous neural network acceleration works~\cite{ishida2020supernpu} targeting RSFQ superconducting logic, recent research~\cite{10040729} proposes a crossbar synapse array architecture designed for implementing BNN models tailored for AQFP logic.
However, unresolved issues like current attenuation, limited scalability, and the randomized behavior of AQFP buffers still hinder the true implementation of AQFP-based crossbar array architecture. Our proposed framework addresses and resolves these challenges, making it a feasible solution.}

\subsection{Stochastic Computing}
\label{sec:SC_basic}

\begin{figure}[t]
  \centering
  \includegraphics[width=1\linewidth]{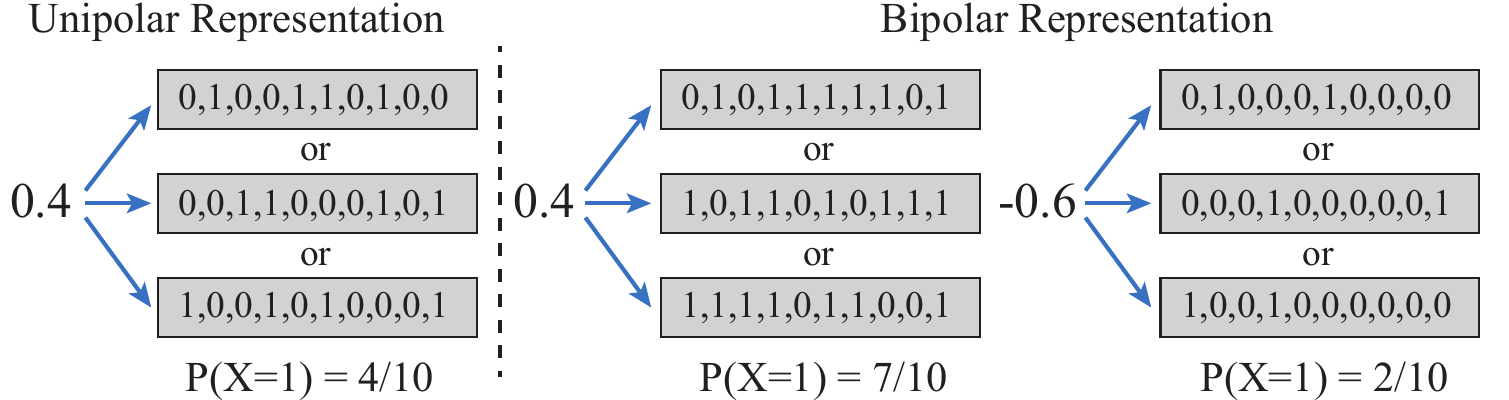}
  % \vspace{-1em}
  \caption{Examples of the unipolar and bipolar representations of stochastic numbers.}
  \label{fig:stochastic_numbers}
  % \vspace{-1em}
\end{figure}

Stochastic computing (SC) is a paradigm that represents a number, named stochastic number (SN), by counting the number of ones in a bitstream. For example, the bitstream 0100110100, containing four ones in a 10-bit stream, represents a real number ${x=P_X=4/10=0.4}$. (Here we use $X$ to represent the stochastic bitstream, whereas $x$ represents the real value associated with $X$.) 
In the bit-stream, each bit is independent and identically distributed (i.i.d.). 
In addition to the above unipolar encoding format, SC can also represent numbers in the range of [-1, 1] using the bipolar encoding format. 
Concretely, a real number $x$ is processed by $P(X=1)=(x+1)/2$. 
Hence, 0.4 can be represented by 1011011101, as $P(X=1) = (0.4+1)/2 = 7/10$. 
$-0.6$ can be represented by 0100100000, with $P(X=1) = (-0.6+1)/2 = 2/10$.
Figure~\ref{fig:stochastic_numbers} shows the examples of different SN representation format.

\rev{Recent work SC-AQFP~\cite{cai2019stochastic} develops AQFP-based DNN acceleration framework trying to use stochastic computing to realize the whole DNN implementation. But it can only work on a very small network for simple tasks (e.g., MNIST) without complex layers (e.g., batch normalization) and requires a pretty large bit-stream length (i.e., 256$\sim$2048). Compared with SC-AQFP, our proposed~\M~contributes a new computational paradigm, where stochastic computing is used as a component for the accumulation of intermediate results, which can work on larger DNN and requires smaller bit-stream length (i.e., 16$\sim$32).}

% \textcolor{red}{Stochastic Computing related work}

\section{Challenges and Motivations}

\rev{As mentioned in the introduction, the characteristics of the AQFP buffer well match the needs of computation in BNN models. It is appealing to design the ultra energy-efficient AQFP-based BNN accelerator. Recent work~\cite{10040729} proposes an AQFP-based crossbar synapse array architecture targeting BNN model implementation. This architecture pre-stores BNN weights and deploys XNOR macro inside logic-in-memory cells, which can achieve energy-efficient in-memory computing theoretically. But the randomized behavior of AQFP buffers, current attenuation within the crossbar, the limited hardware scalability, and the hardware configuration problem makes it hard to realize a practical deployment.}

\rev{\textbf{Randomized Behavior of AQFP Buffer:}
Because of the thermal noise and/or quantum fluctuation impact, the output of AQFP buffer presents randomized switching behavior, especially when input current amplitude falls in a certain range, known as a finite ``gray-zone" $\Delta I_{in}$ \cite{filippov1995}, in which the AQFP buffer may not be able to precisely detect the direction of the input current. Such a phenomenon introduces in-accuracy in BNN computation and may lead to a degraded network accuracy eventually.}

\rev{To handle this problem, we first investigate the randomized behavior and simulate this phenomenon within our research scope (4.2K), then incorporate it into our proposed AQFP-aware BNN training algorithm. (Section~\ref{sec:current_attenuation_analysis} and Section~\ref{sec:training})
}

\revm{Previous ReRAM and PCM-based work~\cite{kariyappa2021noise} also consider randomness on devices. There are two types of noises considered. Programming Noise and Draft Noise. People usually add a random variable to the original weights to mimic the potential noise/imprecision when mapping the model on different products/hardware, and make a trained model have overall better performance/robustness on different products/hardware. These noises are deterministic after a model is mapped to a specific hardware. They are not data-dependent. On the contrary, the randomized behavior in AQFP devices is data-dependent, which depends on both $I_in$ and hardware configuration for each computation. Therefore, we need to analyze the probability of the intermediate results and incorporate this randomized behavior inside the training algorithm.}

\rev{\textbf{Crossbar Current Attenuation and Scalability Problem:}
When building an AQFP-based crossbar, the accumulated current in the analog domain on each crossbar column will suffer from the current attenuation caused by the increasing superconductive inductance. The relationship between accumulated current amplitude with the mathematical value (the latent activation value in BNN) varies in terms of the crossbar size which increases the randomized behavior in the value domain because the attenuated input current amplitude is more likely to fall into the ``gray-zone" of the AQFP buffer. 
As a result, such randomness in the value domain is intensified when the crossbar size becomes larger. Since excessive current attenuation results in completely randomized output, the AQFP crossbar scalability is limited and it is not able to accommodate all the weights from a BNN layer or a convolutional filter. To overcome the limited scalability of AQFP crossbar, we use multiple crossbars to accumulate the intermediate result of each BNN filter. To mitigate the impact of the current attenuation on BNN computation, we investigate the impact of crossbar size, formulate the current amplitude into the value used in BNN computation, and incorporate the factor of current attenuation into the AQFP-aware BNN training (Section~\ref{sec:current_attenuation_analysis}).}

\rev{\textbf{Accumulation of Intermediate Result Problem:}}
\rev{In the design of~\cite{10040729}, AQFP buffer is used as the neuron circuit of the crossbar (Section~\ref{sec:crossbar}), which functions both a sign operator and an analog-digital-converter (ADC), directly outputting the 1-bit binarized results. This architecture is ultra-energy-efficient but requires one column of the crossbar to contain a whole filter computation in BNN. But as mentioned above, the limited scalability of AQFP crossbar may not satisfy the demand of BNN model and we need to use multiple crossbars to do the intermediate results' accumulation.}

\rev{To handle the problem of intermediate results' accumulation as well as preserve the accuracy impacted by the AQFP buffer randomized behavior, we design a novel SC-based accumulation module circuit as the output peripheral circuit to add up the intermediate results from each crossbar and convert the stochastic numbers back to 1-bit value as the input of the next layer (Section~\ref{sec:sc_accumulator}). }

\rev{\textbf{Hardware Configuration Problem:}
In general, the crossbar accelerator designs prefer a larger crossbar size and coarse-grained computations to ensure a higher computation throughput and energy efficiency~\cite{shafiee2016isaac,ankit2019puma,li2020timely}. The AQFP-based design becomes more complex since the hardware configurations, such as the crossbar size, the ``gray-zone'' width, the threshold current of the AQFP comparator (illustrated in Section~\ref{sec:current_attenuation_analysis}), and the bit-stream length of SNs not only affect the energy efficiency but also affect the randomized behavior as well as the model accuracy. We need to make optimization to deliver the best accuracy while considering the hardware constraints.
Therefore, we propose a comprehensive algorithm-hardware co-optimization for both randomized BNN training and hardware configurations (Section~\ref{sec:Configuration Optimization}). To fully leverage the potential of AQFP devices, we also introduce a batch normalization matching method to address the floating-point computation problem induced by BN layer with no additional peripheral circuits (Section~\ref{sec:Configuration Optimization}).}

\section{Hardware Design of AQFP-based Randomized BNN Accelerator} \label{sec:hardware}

In this section, 
we first revisit the AQFP-based crossbar synapse array and the corresponding neuron circuit design proposed in~\cite{10040729} (Section~\ref{sec:crossbar}).
% and Section~\ref{sec:neuron_circuit}).
% we will introduce the hardware design of our AQFP-based randomized BNN accelerator. In Section~\ref{sec:crossbar}, we introduce our AQFP-based crossbar synapse array design, which is used to compute the binary matrix multiplication between weights and input feature maps/activations in BNNs.
% The AQFP-based neuron circuit design is introduced in Section~\ref{sec:neuron_circuit}, which works as the ADCs to convert the accumulated current from analog domain to the logic states.
Then, we make a comprehensive analysis of the randomized behavior of AQFP buffer and crossbar current attenuation and propose our novel designs.
In Section~\ref{sec:current_attenuation_analysis}, we explore the impact of crossbar size on the current attenuation and analysis the randomized behavior of AQFP buffer and formulate the current amplitude into the value used in BNN computation.
In Section~\ref{sec:sc_accumulator}, we propose a stochastic computing-based accumulation module to accumulate the intermediate computation results from the crossbar columns by considering the randomized outputs from the neuron circuits.

\subsection{AQFP-based Crossbar Synapse Array Design for BNN}
\label{sec:crossbar}
\begin{figure}[t]
  \centering
  \includegraphics[width=0.8\linewidth]{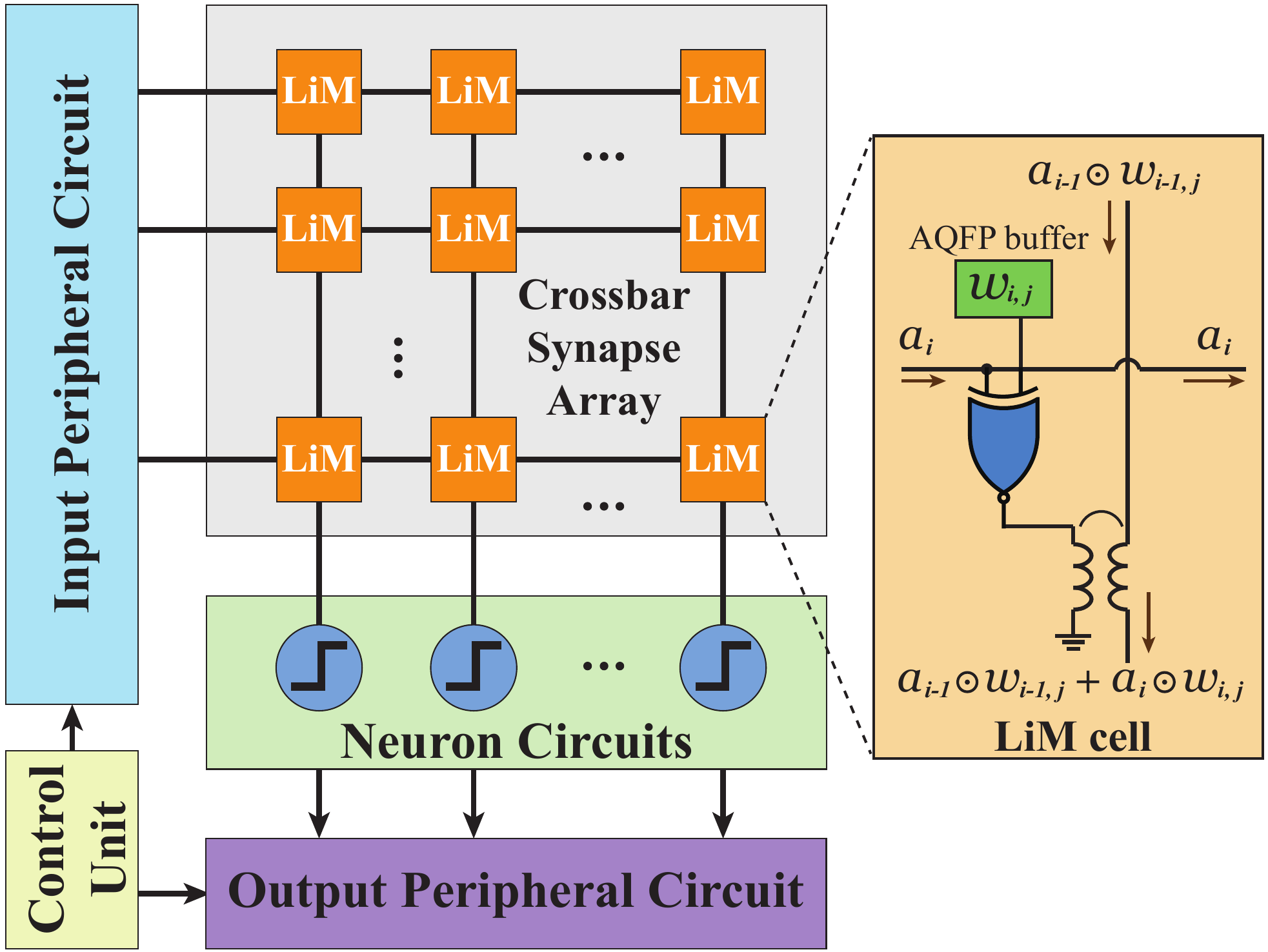}
  \caption{AQFP-based crossbar synapse array circuit architecture.}
  % \vspace{-1em}
  \label{fig:crossbar}
\end{figure}

Although BNNs employ binary weights and activations, they still suffer from significant data movement between the memory and computing units in conventional Von Neumann architectures. This data movement can lead to performance bottlenecks and increased energy consumption.
Considering the nature of the AQFP buffer that can be used as a single-bit memory cell and its output current can be easily accumulated in the analog domain,
\rev{the logic-in-memory (LiM) array-based architecture is used to perform BNNs, employing the in-memory/near-memory computing concept.} 
Figure \ref{fig:crossbar} illustrates the circuit architecture of the AQFP BNNs. 
The binarized weights are pre-stored in the 1-bit AQFP LiM cells and multiplied by the in-cell XNOR macro. 
The output of each LiM cell is the multiplication result of the input activation $a_i$ in the $i_{th}$ row of the crossbar and the corresponding pre-stored weight $w_{i,j}$ in the $i_{th}$ row and $j_{th}$ column of the synapse array, and the multiplication result is represneted as $a_i\bigodot w_{i,j}$.
\rev{Being different from conventional popcount-based accumulation in BNNs, the crossbar adopts an analog summation approach to add up all the outputs directly since the logic `1' and `0' in AQFP are represented by positive and negative current pulses. 
The accumulation result represented by the current sum-up of each column in the illustrated synapse array will be sent to the neuron circuits.}

As shown in Figure~\ref{fig:AQFP_basic}, a basic AQFP gate consists of two inductor-Josephson-junction loops $L_1$-$J_1$ and $L_2$-$J_2$, and the output logic state is denoted by the positive or negative current flowing through the output inductor $L_{out}$. 
Therefore, an AQFP buffer can also serve as a current sensor since AQPF buffers can detect directions/signs of the input current and convert them into `0's or `1's.
This unique characteristic makes AQFP naturally able to be used as the analog-digital-converters (ADCs) in BNN since the BNN also requires 1-bit representation for the intermediate computation results such as the output feature maps.

Therefore, the neuron circuit can be simply built by using the AQFP buffers.
For a specific column of the crossbar, the output currents of each LiM cell are merged by the magnetic coupling and obtain the accumulated current.
Then, depending on the direction of the accumulated current, an AQFP buffer serves as both a sign function and an ADC to binarize and covert the accumulated current into logic state `0' or `1'.

% \subsection{Stochastic Characterization of AQFP Unit}
\subsection{Randomized Behavior of AQFP buffer and Crossbar Current Attenuation Analysis}
\label{sec:current_attenuation_analysis}

Ideally, the neuron circuit in the BNN desires to generate deterministic results to ensure accurate computation. 
The randomized behavior of AQFP-based neuron circuits may introduce computation inaccuracy, leading to a potential accuracy drop eventually.
Therefore, we need an effective way to quantify the randomized behavior, so that we can integrate it into the BNN training process. And with such a randomized-aware trained BNN, the accuracy can be significantly preserved.
Moreover, understanding the relationship between the crossbar size and randomized behavior can also help us select appropriate hardware configurations for the implementation.

\begin{figure}[b]
  \centering
  % \vspace{-1em}
\includegraphics[width=0.8\linewidth]{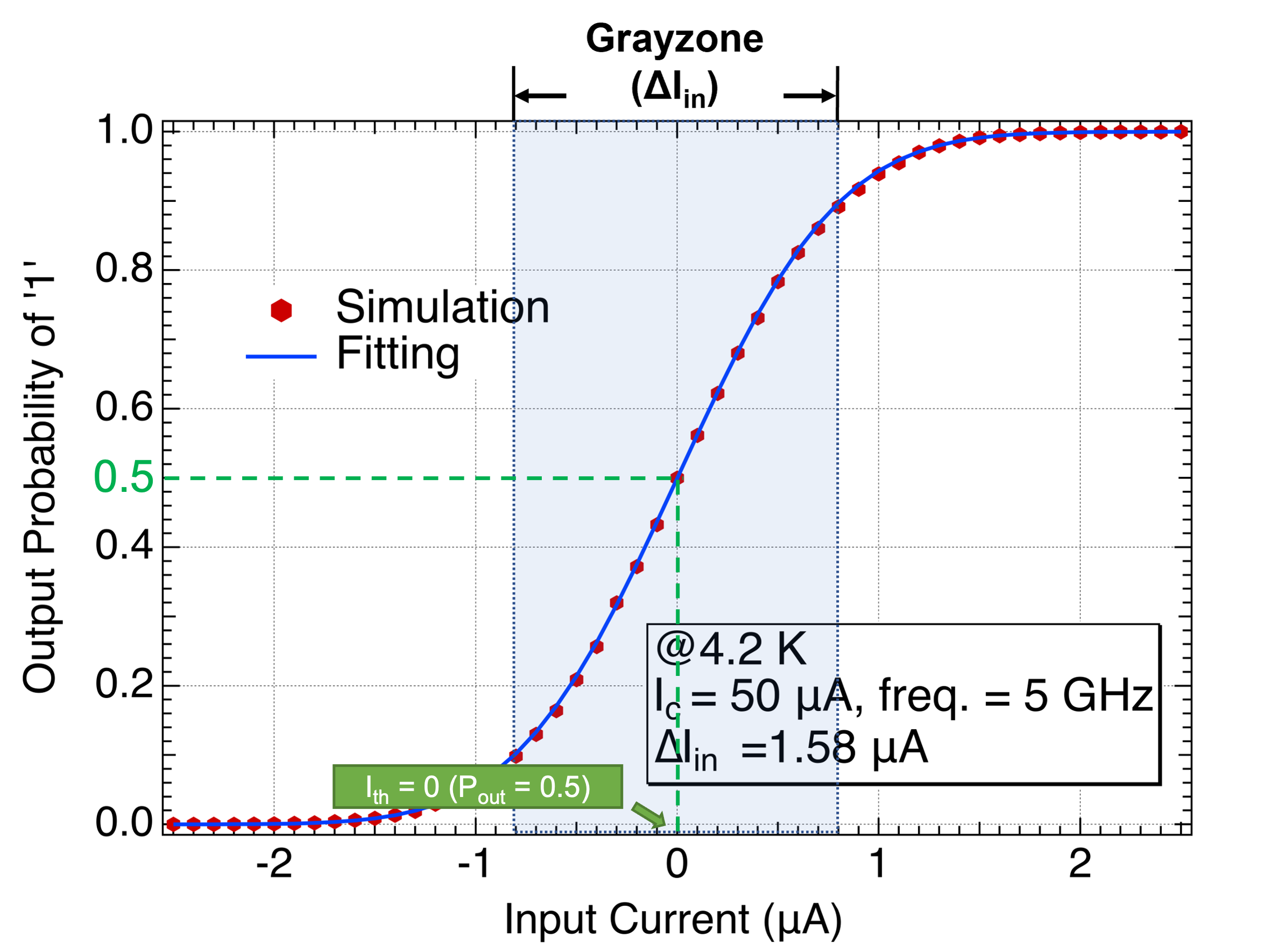}
% \vspace{-1em}
  \caption{The relationship between output probability of "1" with input current on AQFP buffer.}
  \label{fig:Probability}
  % \vspace{-1em}
\end{figure}

As we mentioned earlier, the AQFP buffer can serve as a current sensor to detect directions/signs of the input current.
However, randomized switching behavior exists in an AQFP comparator when input current amplitude falls in a certain range, known as a finite ``gray-zone" $\Delta I_{in}$ \cite{filippov1995}, resulting in a finite output probability $0<P(I_{in})<1$, introduced by the thermal or quantum fluctuation.
Quantitative research \cite{walls_quantum_2002} on the quantum fluctuation effect on Josephson device shows that $\Delta I_{in}$ grows at high temperatures due to thermal noise, whereas at $T\rightarrow0$, it saturates due to quantum fluctuations. Within our research scope (4.2K), we only consider thermal fluctuations as noise sources.
Figure \ref{fig:Probability} shows the output probability of `1' corresponding to a given input current amplitude in micro-ampere level, where the boundary of randomized switching is around $\pm\SI{2}{\micro\A}$.

The probability of output of a forward current from AQFP buffer can be formulated as:
\begin{equation}
P\left(I_{i n}\right)=0.5+0.5 \operatorname{erf}\left(\sqrt{\pi} \frac{\left(I_{i n}-I_{t h}\right)}{\Delta I_{i n}}\right),
\label{equ:probability}
\end{equation}
where $I_{in}$ is the input current amplitude of the AQFP buffer, which is accumulated through the whole column in the crossbar synapse array. $\Delta I_{in}$ means the width of the ``gray-zone''. $I_{th}$ means the current threshold which can be adjusted manually, and $\operatorname{erf}\left(\cdot\right)$ is the error function.

To better explore the impact of this randomized behavior on the BNN and quantify it, we conduct an analysis of the crossbar current attenuation.

\begin{figure}[t]
  \centering
  \includegraphics[width=0.9\linewidth]{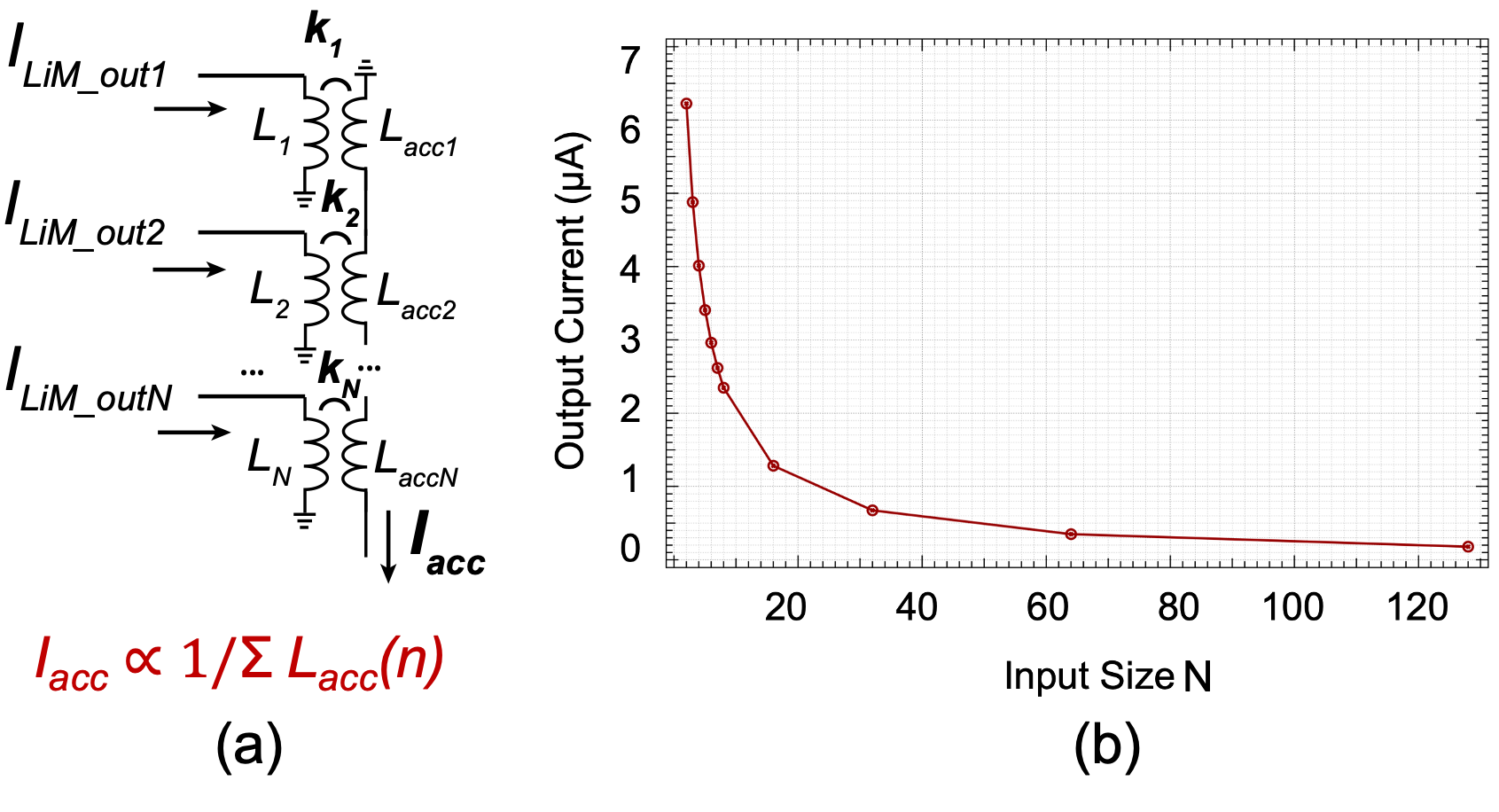}
  % \vspace{-1em}
  \caption{(a) Schematic of analog accumulation circuit. (b) Current Attenuation Curve. The relationship between output current with crossbar synapse array size.}
  \label{fig:current-value}
  % \vspace{-1em}
\end{figure}

For the input of the crossbar synapse array, we use $+70\mu A$ and $-70\mu A$ to present value of $+1$ and $-1$, respectively. Since the current is added together (merged) in an analog manner via superconductive inductance, the merged current amplitude inevitably attenuates as more inputs in the merging circuits bring larger inductance.
As shown in Fig.~\ref{fig:current-value}, we measure the degree of current attenuation under different crossbar synapse array sizes.
According to the rationale of the current attenuation, it is reasonable to see the amplitude of the output current decrease as the crossbar size increases. 
Then, we generate a corresponding mathematical fitting curve of it. The curve can be expressed in the form of:

\begin{equation}
I_{1}(C_s)=A \cdot C_s^{-B},
\end{equation}
where, $I_1$ is the output current amplitude representing the value of 1, $C_s$ is the size of crossbar synapse array, $A$ and $B$ are positive constants of fitting parameters.
 
In consequence, the current amplitude representing the logic state of `1' in the neural network varies according to the size of the crossbar synapse array.
We need to figure out the relationship between the output current amplitude with the presented value and convert the current amplitude to the specific value in the intermediate feature map of the neural network.

% \subsection{Current Attenuation Analysis}

% With our AQFP-based computing unit design for binary neural networks, we need to convert the current amplitude to the specific value in the intermediate feature map of the neural network.

% \textcolor{red}{Need: The reason for Current Attenuation.}

To this end, we can convert the probability Equation~(\ref{equ:probability}) into the DNN value version:

% \begin{equation}
% P\left(I_{i n}\right)=0.5+0.5 \operatorname{erf}\left(\sqrt{\pi} \frac{\left(I_{i n}-I_{t h}\right)/I_{1}}{\Delta I_{i n}/I_1}\right)
% \end{equation}

\begin{equation}
P_v\left(V_{i n}\right)=0.5+0.5 \operatorname{erf}\left(\sqrt{\pi} \frac{\left(V_{i n}-V_{t h}\right)}{\Delta V_{i n}(C_s)}\right),
\label{equ:probability-value}
\end{equation}
where, $V_{in}$ is the mathematical value converted from the input current of AQFP buffer, $V_{th}$ and $\Delta V_{in}(C_s)$ are the counterpart of $I_{th}$ and $\Delta I_{in}$, respectively. $\Delta V_{in}(C_s)$ can be presented as follows:

\begin{equation}
\Delta V_{in}(C_s)=\Delta I_{i n}/I_1(C_s).
\end{equation}

With the DNN value version of the probability expression, we make it possible to consider the AQFP randomized behavior in the BNN training process.

\subsection{Stochastic Computing-based Accumulation Module Design}
\label{sec:sc_accumulator}

Though the randomized behavior of AQFP buffer is not an ideal property for the neuron circuit design, it also makes the AQFP buffer inherently compatible with the stochastic computing (SC) technique.
In SC, the stochastic number (SN) is used to represent the value of a number, which consists of a time-independent bit sequence (as introduced in Section~\ref{sec:SC_basic}).
Since the randomized behavior that appears in the AQPF buffer presents an output probability dependence on the input current amplitude, which can provide a sufficient level of SNs through a certain observation window with almost no hardware overhead.

\begin{figure}[t]
  \centering
\includegraphics[width=0.8\linewidth]{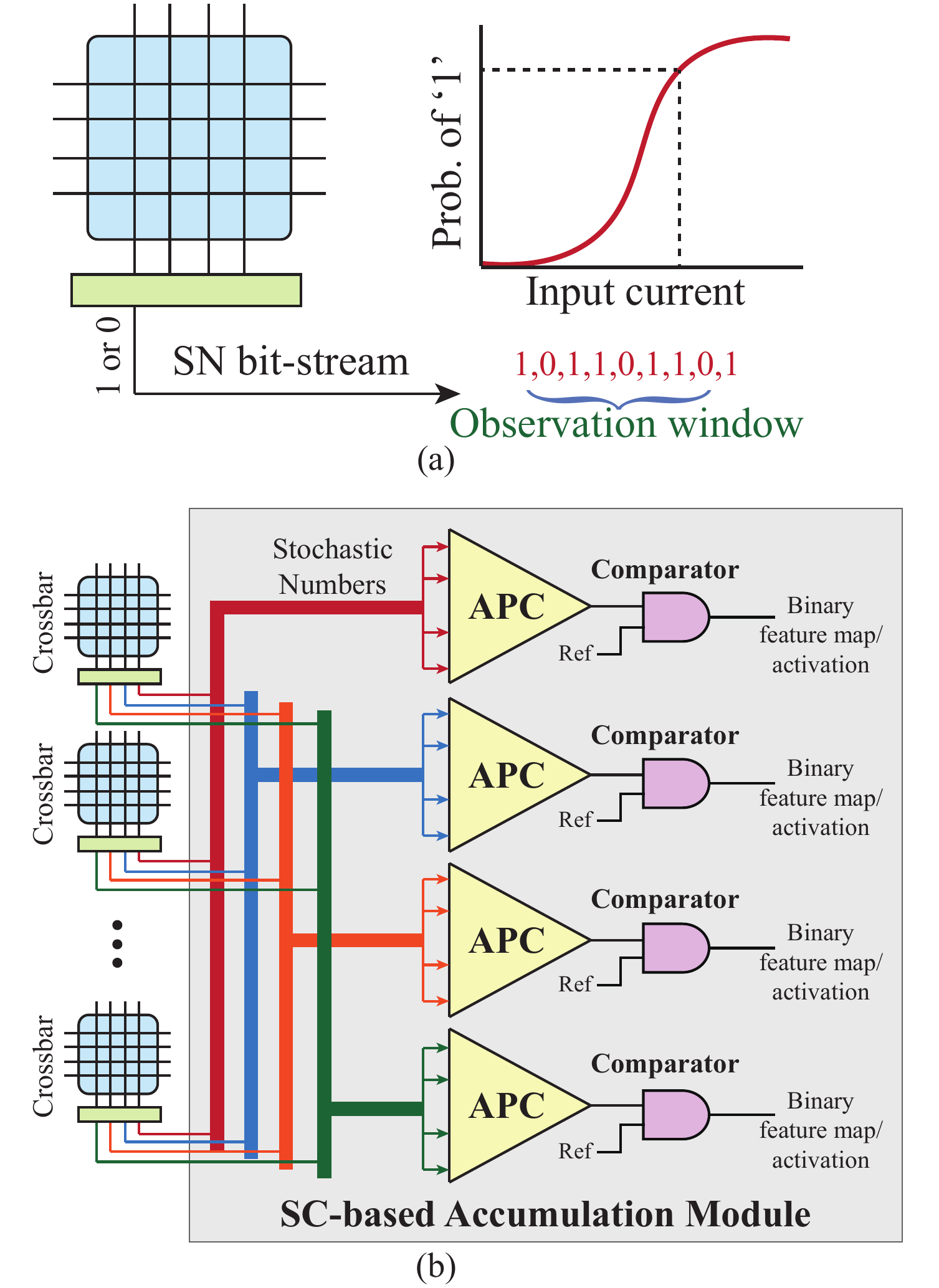}
  % \vspace{-1em}
  \caption{(a) Convert intermediate results to stochastic number through a certrain observation window. (b) Architecture design of stochastic computing-based accumulation module.}
  \label{fig:sc_accumulator}
  % \vspace{-1.2em}
\end{figure}

For example, as shown in Figure~\ref{fig:sc_accumulator} (a), for each clock cycle/phase, an AQFP buffer in the neuron circuit will generate a 1-bit output with the probability of `1' or `0' depending on the accumulated current from the corresponding crossbar column.
Using the 1-bit output directly carries a higher risk of being affected by the randomized behavior of the AQFP buffer.
But, if we allow a longer observation window for the output of the neuron circuit while keeping the input of the crossbar unchanged, we can obtain an output bit-stream, which is naturally a stochastic number, thanks to the true randomness property of the AQFP buffer~\cite{gentle2003random,takeuchi2013measurement}.

Limited by the crossbar current attenuation property and the hardware manufacture constraints, the crossbar size cannot be arbitrarily large.
Therefore, multiple crossbars are needed to accommodate all the weights of the same BNN layer.
To make the convolution computation correct, it is required to accumulate the SNs from the same column of different crossbars.
And we propose our SC-based accumulation module for the inter-crossbar accumulation.
As shown in Figure~\ref{fig:sc_accumulator} (b),
we choose to use the approximate parallel counters (APCs)~\cite{kim2015approximate} to perform the SN accumulation between different crossbars.
The APC counts the number of 1s in the inputs and represents
the result with a binary number. 
This method consumes fewer logic gates compared with the conventional accumulative parallel counter~\cite{parhami1995accumulative,kim2015approximate}.
A comparator is followed by the APC to perform as a step function to generate 1-bit feature map/activation of BNN.
Note that all the logic cells/circuits, such as APCs and comparators are designed based on the AQFP standard cell library consisting of all the AQFP logic gates including AND, OR, buffer, inverter, majority, splitter and read-out interfaces. 
The binary feature maps and activations are represented by positive and negative currents so that they can be directly used as the input for the crossbars for the next level computation.

In general, the larger SN length will result in a higher accuracy of SC, but at the cost of longer computation clock cycles/phases.
In our work, we also make the SC bit-stream length one of the dimensions in our algorithm-hardware co-optimization (more details in Section~\ref{sec:ame}).
By incorporating SC and using our SC-based accumulation module, the possible accuracy loss introduced by the AQFP neuron circuit can be efficiently and effectively resolved.

\begin{figure}[t]
  \centering
\includegraphics[width=0.9\linewidth]{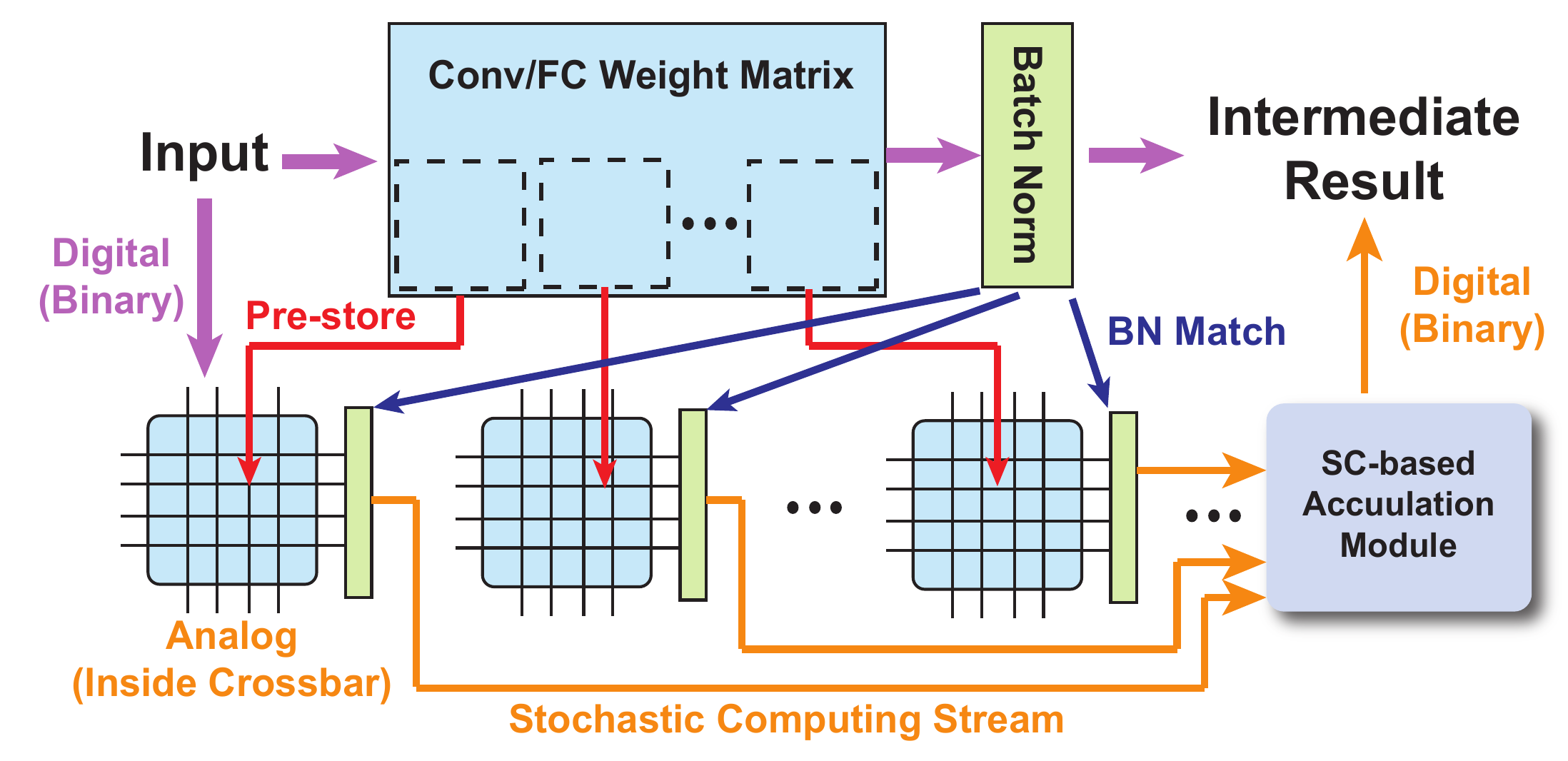}
  % \vspace{-1em}
  \caption{\revm{Computation matching from software model (upside) to hardware architecture (downside).}}
  \label{fig:soft-hard-match}
  % \vspace{-1.2em}
\end{figure}

\revm{To have a better understanding of  where each BNN computational block is implemented, we show the overall matching graph as Fig.~\ref{fig:soft-hard-match}. The weight matrices in BNN blocks are separated and pre-stored in each AQFP crossbar. The batch normalization is directly converted and matched into the neuron circuits after each crossbar without additional cost (refer to Section~\ref{sec:BN}). SC-based accumulation module is used to collect the output of each neuron circuit and generate the intermediate result of each BNN block. The data representations are marked for the corresponding data flows, e.g., Analog, stochastic computing stream, and digital.}

\subsection{Clocking Scheme Adjustment-based Circuit Optimization}
\label{sec: clock-scheme}
In AQFP, all logic gates are synchronized by a multi-phase clock, facilitating data propagation between adjacent logic stages during a sufficient overlapping window of their respective clock phases. This distinctive characteristic necessitates a minimum of a 3-phase clock system. Current AQFP designs commonly employ a 4-phase clocking system, as it simplifies the testing process: a 4-phase clock can be easily generated using a 2-phase ac with a 90-degree phase difference in conjunction with a dc offset. Due to the synchronization nature of AQFP, numerous buffers must be inserted to ensure that all logic paths are balanced, preventing possible data propagation failure caused by the non-overlap of adjacent logic stages in a typical 4-phase clocking scheme. However, increasing the clock phase for the computing part can significantly reduce the buffers required for path-balancing \cite{saitotas2021}. This is because the clock phase overlap exists not only in adjacent logic stages but also in non-adjacent stages. Our simulations indicate that the total Josephson junction (JJ) count can be reduced by at least 20.8\% and 27.3\%, assuming 8-phase and 16-phase clocking, respectively. On the other hand, the buffer-chain-based memory (BCM) employed in this study is achieved in a fully balanced structure without any inserted buffer, and the clock is independent of the computing part. Thus, we propose an alternative approach that involves reducing the number of clock phases in the memory design from the original 4 phases to 3 phases, resulting in a 20\% reduction in the total JJ count of the memory component. These simulations demonstrate the significant potential for clock phase adjustment-based component circuit optimization in enhancing the performance and efficiency of AQFP-based computing systems.
\section{Algorithm and Hardware \\
Co-optimization} 
\label{sec:design}

\subsection{Randomized-aware Binary Neural Network Training}
\label{sec:training}

As the existence of the randomized behavior of AQFP buffer, training a BNN normally will lead to a significant performance mismatch between the pure software results and actual implementation on hardware, resulting in severe accuracy degradation.
To mitigate this issue, it is desirable to make the training process of BNN randomized-aware.

Given a DNN, for ease of representation, we simply denote its per-layer real-valued weights as $w_r$ and the inputs as $a_r$. Then, the convolutional result can be expressed as:

\begin{equation}
Y=\operatorname{CONV}\left(a_r, w_r\right),
\end{equation}
where $\operatorname{CONV}\left(\cdot \right)$ represents the standard convolution. For simplicity, we omit the effect of stochastic computing and non-linear operation in this subsection.

Binarized quantization aims to quantizes weight $\mathcal{w}_r$ and activation $\mathcal{a}_r$ to binarized levels, i.e., $\{+1, -1\}$. 
Following XNORNet~\cite{rastegari2016xnor}, given $x_r$, the corresponding binarized value $x_b$ can be achieved by the sign function:

\begin{equation}
x_b=\operatorname{sign}\left(x_r\right)= \begin{cases}+1, & \text { if } x_r \geq 0, \\ -1, & \text {otherwise},\end{cases}
\label{equ:bnnx}
\end{equation}

Taking the AQFP randomized behavior into consideration, each activation value is generated with the value probability function. Different from the conventional BNN quantization, the randomized activation $a_b$ can be presented as:
% Substituting Equation~(\ref{equ:probability-value}) into Equation~(\ref{equ:bnnx}), the binarized weight $w_b$ and activation $a_b$ can be presented as:

% \begin{equation}
% w_b=\operatorname{sign}\left(w_r\right)= \begin{cases}+1, & \text { if } w_r \geq 0, \\ -1, & \text { otherwise },\end{cases}
% \end{equation}

\begin{equation}
a_b=\operatorname{sign}\left(a_r\right)= \begin{cases}+1, & \text {with probability} ~P_v\left(a_r\right), \\ -1, & \text {with probability} ~1 - P_v\left(a_r\right),\end{cases}
\label{equ:bnn}
\end{equation}

To mitigate the large quantization error in 
DNN binarization, XNOR-Net~\cite{rastegari2016xnor} applies two scaling factors for the quantized weights $w_b$ and activations $a_b$, respectively. Since weight and activation are multiplied in convolution layers, we can simplify these two scaling factors as one parameter, denoted as $\alpha$. Then, the binary
convolution operation can be formulated as:

\begin{equation}
Y_b = \operatorname{BCONV}\left(a_b, w_b\right) \odot \alpha,
\label{equ:forward}
\end{equation}
where $\operatorname{BCONV}\left(\cdot\right)$
denotes the binary convolution which includes bit-wise operations XNOR. $\odot$ represents the element-wise multiplication. Here $\alpha$ is set to be a learnable vector that contains independent values for each output channel.

% Then, following~\cite{xu2021recu}, the quantization error of BNN can be defined as:

% \begin{equation}
% \mathrm{QE}=\int_{-\infty}^{+\infty} f\left(w_r\right)\left(w_r-w_b\right)^2 \mathrm{~d} w_r,
% \end{equation}
% where $f\left(w_r\right)$ is the probability density function of $w_r$. 

For BNN training, the forward-propagation is expressed by Equation~(\ref{equ:forward}) with the binarized values $w_b$ and $a_b$, while the real-valued $w_r$ and $a_r$ are updated during the back-propagation.
However, the gradient of the sign function is an impulse function that breaks the transitivity of the derivative. The back-propagation can not be processed directly. 
Following STE~\cite{bengio2013estimating}, we compute the approximate gradient of the 
loss function $L$, as following:

\begin{equation}
\frac{\partial L}{\partial w_r}=\frac{\partial L}{\partial w_b} \cdot \frac{\partial w_b}{\partial w_r} \approx \frac{\partial L}{\partial w_b},
\end{equation}

For the gradient of the activations, since the AQFP probability function has already turned the sign function into the error function, we can leverage this characteristic to achieve the back-propagation instead of using piece-wise polynomial function~\cite{liu2018bi}. Using the expected value of $a_b$ as the approximation, we can have the AQFP randomized-aware back-propagation as follows:

\begin{equation}
\frac{\partial L}{\partial a_r}=\frac{\partial L}{\partial a_b} \cdot \frac{\partial a_b}{\partial a_r} \approx \frac{\partial L}{\partial a_b} \cdot \frac{\partial \mathbb{E}\left(a_b\right)}{\partial a_r},
\end{equation}
where $\mathbb{E}\left(a_b\right)=\operatorname{erf}\left(\sqrt{\pi} \frac{\left(a_r-V_{t h}\right)}{\Delta V_{i n}(C_s)}\right)$ 
% is an error function which can be presented as.

% The partial derivative is equal to:

% \begin{equation}
% \begin{aligned}
% \frac{\partial \mathbb{E}\left(a_b\right)}{\partial a_r} & =\frac{\partial \operatorname{erf}\left(\sqrt{\pi} \frac{\left(a_r-V_{t h}\right)}{\Delta V_{i n}(C_s)}\right)}{\partial a_r} \\
% & = \frac{\partial \sqrt{\pi} \frac{\left(a_r-V_{t h}\right)}{\Delta V_{i n}(C_s)}}{\partial a_r} \cdot \frac{2}{\sqrt{\pi}} e^{-\left(\sqrt{\pi}\frac{\left(a_r-V_{t h}\right)}{\Delta V_{i n}(C_s)}\right)^2}\\
% & = \left(2 e^{-\frac{\pi a_r^2-2 \pi V_{t h} \cdot a_r+\pi V_{t h}^2}{(\Delta V_{i n}(C_s))^2}}\right)/\Delta V_{i n}(C_s).
% \end{aligned}
% \end{equation}
% (Replaced with description)

In this way, we implement both forward-propagation and backward-propagation, which achieves the AQFP randomized-aware training.

\subsection{Batch Normalization Matching} \label{sec:BN}
% \zg{From now on, we drop the superscript “r” for real-valued weights for simplicity.}

Batch Normalization (BN)~\cite{ioffe2015batch} is a DNN layer that normalizes the activation values in the mini-batch during training. Many neural networks use the BN since it is important in stabilizing and accelerating the training process. But BN brings additional floating-point computation in the inference period which causes inefficiency in the BNN implementation on AQFP devices. In this section, we propose an AQFP-aware BN matching technique. 

BN can be described by the following equation:

\begin{equation}
Y=\gamma \frac{X-\mu}{\sqrt{\sigma^2+\epsilon}}+\beta
\label{equ:bn}
\end{equation}
where $X$ and $Y$ are the input and output of BN, and
$\gamma$, $\beta$, $\mu$, and $\sigma$ stand for weight, bias, mean, and standard deviation, respectively. $\epsilon$ is a small constant value to prevent the potential zero in the denominator. $\gamma$ and $\beta$ are updated through back-propagation in the training process. $\mu$ and $\sigma$ are updated using a moving average during training but fixed in inference.
Note that BN in the inference process becomes a linear transformation, which makes it possible to convert BN into a simple addition operation in BNN and match crossbar synapse array.

\begin{figure}[t]
  \centering
  \includegraphics[width=0.75\linewidth]{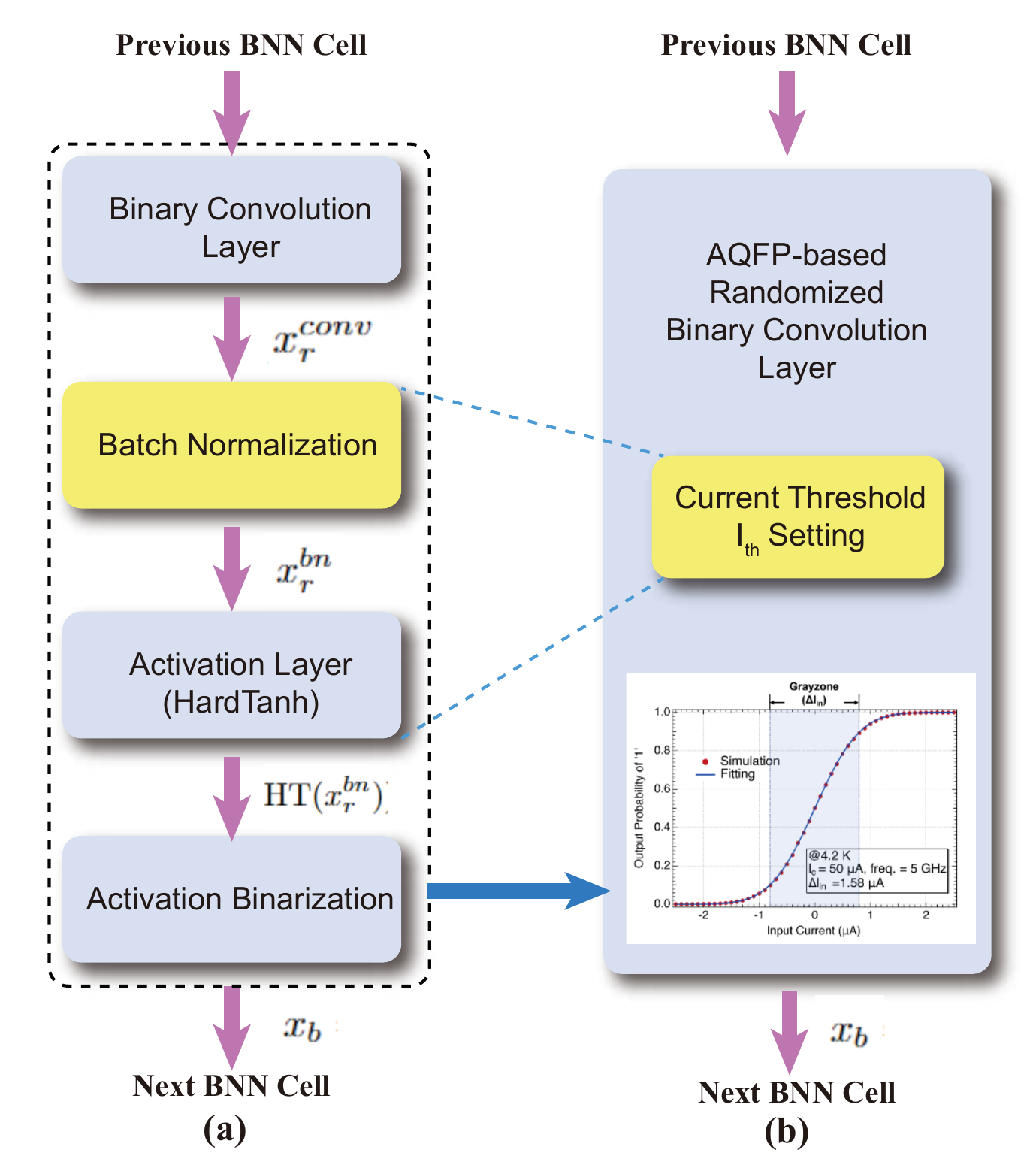}
  \caption{BNN cell architecture. (a) Basic BNN convolution cell, (b) converted AQFP-based randomized BNN convolution cell.}
  % \vspace{-1em}
  \label{fig:DNN structure}
\end{figure}

As shown in Fig.~\ref{fig:DNN structure} (a), we use a common BNN cell as an example. The data get through the convolution layer followed by a BN layer and an activation layer (HardTanh), then input into the binarization layer before getting into the next BNN cell. The transferred values before the binarization layer are deformed back to floating-point values $x_r^{bn}$ due to the existence of BN. Given the output values $x_r^{conv}$ for binary convolution layer, the $x_r^{bn}$ for BN layer, and the scaling factor $\alpha$, the BN can be rewritten as:

\begin{equation}
x_r^{bn}=\gamma \frac{x_r^{conv} \cdot \alpha-\mu}{\sqrt{\sigma^2+\epsilon}}+\beta
\label{equ:bn}
\end{equation}

% The HardTanh can be indicated as:

% \begin{equation}
% \operatorname{HT}(x)= \begin{cases}1 & \text { if } x>1 \\ -1 & \text { if } x<-1 \\ x & \text { otherwise }.\end{cases}
% \label{equ:hardtanh}
% \end{equation}

% Substituting equation (\ref{equ:hardtanh}) into equation (\ref{equ:bnnx}), we have:
The output $x_b$ of the BNN cell can be indicated as:

\begin{equation}
\begin{aligned}
x_b & =\operatorname{sign}(\operatorname{HT}(x_r^{bn})) = \begin{cases}+1, & \text {if} ~x_r \geq 0, \\ -1, & \text {otherwise},\end{cases}
\end{aligned}
\end{equation}
where $HT$ means the activation function HardTanh.

Combined with equation (\ref{equ:bn}) and AQFP probability function, the whole cell can be merged as:

When $\gamma > 0$: 

% \begin{equation}
% x_b= \begin{cases}-1 & \text { if } x_r^{conv}<-\frac{\beta \sqrt{\sigma^2+\epsilon}}{\gamma \cdot \alpha}+\frac{\mu}{\alpha} \\ +1 & \text { if } x_r^{conv} \geq -\frac{\beta \sqrt{\sigma^2+\epsilon}}{\gamma \cdot \alpha}+\frac{\mu}{\alpha}\end{cases}
% \end{equation}
\begin{equation}
x_b= \begin{cases}+1,  \text {with probability}~P_v\left(D\right), \\ -1,  \text {with probability}~1 - P_v\left(D\right),\end{cases}
\end{equation}

When $\gamma < 0$:
% \begin{equation}
% x_b= \begin{cases}-1 & \text { if } x_r^{conv}>-\frac{\beta \sqrt{\sigma^2+\epsilon}}{\gamma \cdot \alpha}+\frac{\mu}{\alpha}, \\ +1 & \text { if } x_r^{conv} \leq -\frac{\beta \sqrt{\sigma^2+\epsilon}}{\gamma \cdot \alpha}+\frac{\mu}{\alpha}\end{cases}.
% \end{equation}
\begin{equation}
x_b= \begin{cases}+1,  \text {with probability}~1 - P_v\left(D\right), \\ -1,  \text {with probability}~P_v\left(D\right),\end{cases}
\end{equation}
where $D = x_r^{conv}+\frac{\beta \sqrt{\sigma^2+\epsilon}}{\gamma \cdot \alpha}-\frac{\mu}{\alpha}$.

Thus, we can achieve a similar activation format as equation (\ref{equ:bnn}) by leveraging the current threshold mentioned in Equation~(\ref{equ:probability}) with the setting:

\begin{equation}
I_{th} = \left(-\frac{\beta \sqrt{\sigma^2+\epsilon}}{\gamma \cdot \alpha}+\frac{\mu}{\alpha}\right) \cdot I_1(C_s).
\end{equation}

As shown in Fig.~\ref{fig:DNN structure} (b), by adjusting the hardware configuration $I_{th}$ in the AQFP probability function, the whole cell is converted into one randomized binary convolution layer without additional peripheral circuits. If the computation needs to be separated into multiple crossbars with stochastic computing as shown in Fig.~\ref{fig:sc_accumulator} (b), we can divide $I_{th}$ evenly and assign them to the corresponding crossbar.

\subsection{Weight Rectified Clamp Method}
\label{sec:rectified}
As pointed out in~\cite{banner2019post,xu2021recu}, the real-valued weights $w_r$ of a quantized network roughly follow the zero-mean Laplace distribution due to their quantization in the forward propagation. Most weights are gathered
around the distribution peak, while many outliers fall into the two tails, far away from the peak.
These outliers adversely affect the training
of a BNN and slow down the convergence when training BNNs. It is because though the magnitudes of the weights are updated during the back-propagation by gradient descent, the chances of changing their signs are extremely small, which limits the representational ability of BNNs ~\cite{xu2021recu}.

To revive these outlier weights and promote the BNN training performance, we apply weight rectified clamp method following ReCU~\cite{xu2021recu}:

\begin{equation}
\operatorname{ReCU}(w_r)=\max \left(\min \left(w_r, Q_{(\tau)}\right), Q_{(1-\tau)}\right),
\end{equation}
where $Q(\tau)$ and $Q(1 - \tau)$ denote the $\tau$ quantile and $(1 - \tau)$ quantile of Weight, respectively.

As proved in ReCU~\cite{xu2021recu}, this technique can move the outlier weights towards
the distribution peak to increase the probability of changing their signs, which decreases the quantization error, as well as promotes the representational ability of BNNs.

\subsection{Hardware Configuration Optimizations} \label{sec:Configuration Optimization}

In this section, we optimize the hardware configurations of AQFP-based randomized BNN accelerator design, including crossbar synapse array size, stochastic computing bit-stream length, and ``gray-zone'' width $\Delta I_{in}$ by comprehensively considering power consumption, energy efficiency, and hardware computing error. The computing error mainly comes from two aspects: the average mismatch error $\mathrm{AME}$ comes from the output expectation bias of AQFP buffer (see Section~\ref{sec:ame}); and the stochastic computing error including SN quantization error and random fluctuation~\cite{qian2010architecture,baker2020bayesian}. For simplicity, we omit the mathematical deduction of the latter error and use a series of accuracy comparative experiments to directly analyze it (see Section~\ref{sec:SC}).

% \subsection{Stochastic Computing Optimization}

% \begin{figure}[h]
%   \centering
%   \includegraphics[width=1.0\linewidth]{Fig/sc_num24_resize.png}
%   \caption{Relationship between SC bit-stream length with model accuracy. VGG-small trained on CIFAR-10 with four different crossbar sizes are deployed in the comparison. The $\Delta I_{in}$ is set to be 2.4 $\mu A$ in this experiment.}
%   \label{fig:sc_num}
% \end{figure}

\subsubsection{Stochastic Computing Bit-stream Length Optimization} \label{sec:SC}

We take advantage of the probabilistic behavior that appears in the AQFP buffer to achieve stochastic computing among multiple crossbar synapse arrays. In this process, the stochastic computing bit-stream length is a critical configuration that has a close relationship with model accuracy, inference latency, and power consumption. Generally, a large bit-stream-length leads to better accuracy but suffers from longer inference latency and more power consumption. 

To choose a proper bit-stream length, we conduct a series of experiments to explore its behavior on model accuracy. 
Our general observation is that, as the SC bit-stream length increases (from 1), the model accuracy is improved significantly at the beginning but the accuracy stabilizes after the bit-stream length reaches 16$\sim$32.
Therefore, using a bit-stream length longer than 32 will not have considerable gain in accuracy.
% As shown in Fig~\ref{fig:sc_num}, we use VGG-small training on CIFAR-10 as an example, four different crossbar sizes are incorporated in the comparison. 
% We observe that, as the SC bit-stream length increases, the model accuracy improves a lot at the beginning, but maintains a stable value after the bit-stream length reaches 16$\sim$32. 
Compared with pure stochastic computing work which generally requires a pretty large bit-stream length, e.g., 512, 1024, to maintain the stability of computation,~\M \ reaps benefit from the low demand of bit-stream length, thus achieving fast computation speed.
Detailed results can be found in Section~\ref{sec:ablation_SC_length}.

\subsubsection{Optimization for Width of Gray-zone $\Delta I_{in}$ and Crossbar Size} \ 
\label{sec:ame}

Generally, given a crossbar size $C_s$, 
for the stochastic computing of bipolar signals, the information carried in a stochastic stream of bits $X$ is $x = (2P(X=1) - 1 ) \cdot C_s = 2P(X) \cdot C_s - \cdot C_s$,
% ~\cite{brown2001stochastic}, 
where $X$ is the stochastic bitstream, and $x$ represents the real value associated with X ($-C_s \leq x \leq +C_s$). Since the AQFP buffer is used to generate the stochastic number with the probability $P(X=1)=P_v(x)$. the expected value of the carried information $y = (2P_v(x) \cdot C_s - C_s) = \operatorname{erf}\left(\sqrt{\pi} \frac{\left(x-V_{t h}\right)}{\Delta V_{i n}(C_s)}\right)\cdot C_s$ does not exactly match the real value $x$. 
% As shown in Fig.~\ref{fig:SC-mismatch}, the x-axis and y-axis represent the real-valued input $x$ and the
% the expected $y$ carried in the output SN of AQFP buffer, respectively.
The nonlinear probability function of AQFP buffer causes an expectation mismatch, which impacts the robustness and accuracy of the model. We show more comparison results in Section~\ref{sec:results}.

Considering the activation value distribution, the average mismatch error $\mathrm{AME}$ can be defined as:

\begin{equation}
\mathrm{AME}=\frac{1}{C_s}\int_{-C_s}^{+C_s} f\left(x|C_s\right)\left(x-y\right)^2 \mathrm{~d} x,
\end{equation}
% \begin{equation}
% \mathrm{AME}=\frac{1}{C_s}\int_{-C_s}^{+C_s} f\left(x|C_s\right)\left(x-C_s 
%  \cdot \operatorname{erf}\left(\sqrt{\pi} \frac{\left(x-V_{t h}\right)}{\Delta V_{i n}(C_s)}\right)\right)^2 \mathrm{~d} x,
% \end{equation}
where $f\left(x|C_s\right)$ is the probability density function of AQFP-buffer input value $x$. Early works~\cite{banner2019post,zhong2020towards} have shown that the real-valued weight and activation for a quantized model roughly follow a Gaussian distribution. Thus, $f\left(x\right)$ can also be approximated as a Gaussian distribution related to $C_s$, i.e., $f\left(x|C_s\right)\sim N\left(C_s \mu, C_s \sigma^2\right)$.

We optimize the related hardware configuration $\Delta I_{in}$ and crossbar size $C_s$ by minimizing $\mathrm{AME}$. Since $C_s$ is highly related to hardware performance, we first constraint $C_s$ to a range that meets the energy efficiency demand, then adjust both $C_s$ and $\Delta I_{in}$ to find the local optimal solution within it. Related comparison experiments are incorporated in Section~\ref{sec:results}.

\section{Experimental Results}\label{sec:results}

In this section, 
% we first figure out the relationship between crossbar size with hardware-related benchmarks, e.g., power consumption, computation latency, and volume of Josephson-junction (JJ).
% Then, 
\rev{we present optimizations of AQFP hyper-parameters along with comparison results with respect to model accuracy, power consumption, energy efficiency, etc. Finally, we perform thorough optimizations on the overall~\M~to construct the AQFP-aware randomized BNN on both MNIST and CIFAR-10 datasets compared with multiple representative works based on different techniques, including CMOS-based DDN~\cite{chen2014dadiannao} and SyncBNN~\cite{fu2022jbnn}, ReRAM-based IMB~\cite{kim2019memory}, RSFQ/ERSFQ-based JBNN~\cite{fu2022jbnn}, and AQFP-based pure stochastic computing work SC-AQFP~\cite{cai2019stochastic}.
} 
% The goal is to prove the efficiency of our end-to-end AQFP-aware framework~\M~ and optimize the model accuracy under different energy efficiency targets. 
% We present comprehensive comparison results among i)  different $\Delta I_{in}$, crossbar synapse array size, stochastic computing bit-stream-length, etc., and ii) other platforms based on traditional CMOS.

\subsection{Experiment Setup}

\begin{figure}[t]
  \centering
  \includegraphics[width=1\linewidth]{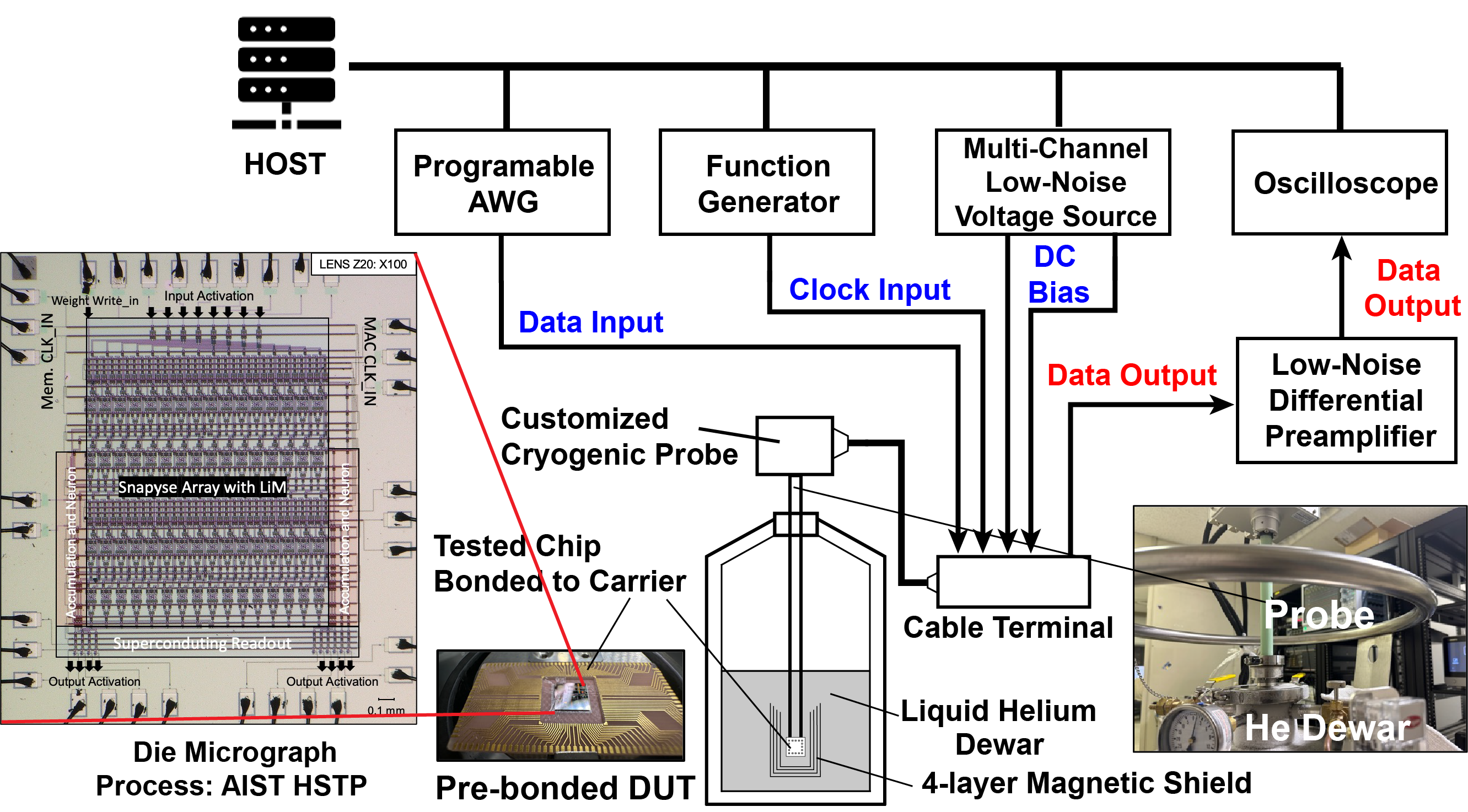}
  % \vspace{-1em}
  \caption{\revm{Module validation setup. Left: Die micrograph of a prototype 8×8 crossbar. Right: Block diagram of the tested system. }}
  \label{fig:tapeout}
  % \vspace{-1em}
\end{figure}

AQFP hardware implementation is achieved using a semi-automated design approach that targets the AIST 4-layer \SI{10}{kA/cm^2} niobium process (HSTP) \cite{hstp}. Analog cells and circuits, such as AQFP neurons and merging circuits (analog accumulation), are manually designed at the Josephson-junction (JJ) level. This process takes into account device characteristics and is optimized with superconductor inductance extraction tools. In contrast, logic cells and circuits, such as logic-in-memory cells, stochastic accumulators (APCs), and comparators, are designed using the AQFP standard cell library. This library consists of all AQFP logic gates, including AND, OR, buffer, inverter, majority, splitter, and read-out interfaces. Figure \ref{fig:tapeout} displays the microphotograph of a fabricated $8\times8$ AQFP crossbar block. The clock/excitation used to drive the entire circuit is a 4-phase sinusoidal current, achieving a \SI{5}{\giga\hertz} clock rate and a \SI{50}{\pico\second} stage-to-stage delay. By introducing a delay-line (micro-stripline) based clocking scheme \cite{delay-clocking}, the overall latency is further reduced. This approach effectively increases the total clock phases to 40 by delaying the sinusoidal current by \SI{5}{\pico\second} between each adjacent logic stage. Circuit-level verification is conducted using a modified version of the Josephson simulator Jsim \cite{Fang1989AJI}, which accounts for thermal noise. \revm{The fabricated $8\times8$ AQFP crossbar block is further validated at \SI{4.2}{\kelvin} inside a liquid helium Dewar, interfaced with a customized cryogenic probe, as illustrated in the right of Figure~\ref{fig:tapeout}, which shows the block diagram of the tested system. The setup includes a chip bonded to a ceramic substrate and housed in a cryogenic probe for testing at 4.2 K. Waveform generators provide data and sinusoidal inputs, while DC voltage sources support a 4-phase clocking scheme. Low-noise differential amplifiers amplify output signals for oscilloscope analysis. A 4-layer shield made by Permalloy effectively blocks the external magnetic field. A remote host manages all input-output operations for automated data handling. Except for the chip and probe, all equipment is at room temperature. Only low-speed module functionality (100 kHz) has been assessed, with high-speed tests planned.}

% \begin{figure}[t]
%   \centering
%   \includegraphics[width=0.95\linewidth]{Fig/bnn_measurement.png}
%   \vspace{-1em}
%   \caption{Module validation setup. (a) Die microphotograph of a prototype 8×8 crossbar. (b) Chip-on-carrier (CoC) configuration. The bare chip is bonded directly onto a ceramic substrate for cryogenic measurement. (c) The chip is mounted inside a cryogenic probe and inserted into a liquid helium dewar for testing at cryogenic temperatures (4.2 K). }
%   \label{fig:tapeout}
%   \vspace{-1em}
% \end{figure}

% \begin{figure}[t]
%   \centering
%   \includegraphics[width=0.65\linewidth]{Fig/liquid helium Dewar.png}
%   \caption{Measuring the AQFP crossbar circuit inside a liquid helium Dewar.}
%   \label{fig:liquid_test}
%   \vspace{-1em}
% \end{figure}

For the thorough optimizations on CIFAR-10 dataset,~\M~trains from scratch and takes 600 epochs to perform the whole AQFP-aware randomized BNN training with a batch size of 256. The learning rate is initialized as 0.1 and decays with a cosine annealing schedule. SGD~\cite{Robbins2007ASA} is used as the optimizer in the training process. Additional training optimizations, such as warmup and weight rectified clamp method are performed during the training. The number of warmup epochs is 5. And we follow the work~\cite{xu2021recu} to initialize the rectified clamp parameter $\tau$ as 0.85, then increase it gradually to the maximum of 0.99 during training. 
% For ImageNet dataset, the total training epoch is 200, and the batch size is 512. Other parameters are the same as the training on CIFAR-10.

\subsection{Hardware Results of the Proposed AQFP-based Crossbar Synapse Array}

Since crossbar synapse array size is a crucial hardware configuration that is highly related to energy efficiency, we first explore the relationship between them.

As shown in Table~\ref{tab:crossbar-array}, we present the hardware results of our proposed AQFP-based crossbar synapse array, including the latency, number of JJs, and energy dissipation (per clock cycle) for one crossbar synapse array of different sizes. 
As the crossbar area increases, all three hardware benchmarks increase but with different growth trends. 
Given a number of JJs, we can get a range of crossbar sizes that meet our energy efficiency requirement.

\begin{table}[htb] 
\centering
% \ra{1.2} % arraystretch
\caption{Circuit latency, JJ count, and energy dissipation under different crossbar sizes.}
\label{tab:statistic}
% \resizebox{0.8\columnwidth}{!}{
% \begin{tabular}{@{}ccc@{}}
\scalebox{0.8}{
\begin{tabular}{c|ccc}
\toprule
\textbf{Crossbar Area} & \textbf{Latency (ps)} & \textbf{\#JJs}  & \textbf{Energy Dissipation (aJ)} \\ \midrule
4$\times$4    &   60     &    384  & 1.92       \\
8$\times$8   &  120     &  1152    &    5.76      \\
16$\times$16   &  240      & 3840       &  19.20   \\
18$\times$18     &  270  & 4752       &  23.76      \\
36$\times$36     &  540  &  17280      &  86.4       \\
72$\times$72     &  1080  &  65664       & 328.32    \\
144$\times$144   &  2160 &  255744    &  1278.72    \\
% c1908                 &             &        &           \\ 
\bottomrule
\end{tabular}}
% \vspace{-1em}
\label{tab:crossbar-array}
\end{table}

\begin{table*}[t] 
\centering
% \ra{1.2} % arraystretch
\caption{Model accuracy on Cifar-10 dataset under different energy efficiency constraints. CMOS-BNN~\cite{knag2020617} has a lower frequency of 13MHz and thus has a relatively higher energy efficiency compared with other CMOS-based work.}
% \resizebox{0.8\columnwidth}{!}{
% \begin{tabular}{@{}ccc@{}}
\scalebox{0.9}{
\begin{tabular}{c|cccccc}
\toprule
\multirow{2}*{\textbf{Design}} &  \multirow{2}*{\textbf{Scheme}} & \multirow{2}*{\textbf{Accuracy}}   & \textbf{Energy Efficiency} & \textbf{Energy Efficiency} & \textbf{Power}  & \textbf{Throughput}  \\ 
& & & \textbf{W/O Cooling (TOPS/W)} & \textbf{W/ Cooling (TOPS/W)} & \textbf{(mW)} &  \textbf{(images/ms)}\\
% \midrule
% \multicolumn{7}{c}{CIFAR-10}
 \midrule
DDN (VGG-Small)~\cite{chen2014dadiannao} & Full-precision & 92.5 & 0.28 & - &  - & -  \\
IMB~\cite{kim2019memory} & Binary   &   87.7     &    82.6  & - & 12.5  & 1.3     \\
\revm{STT-BNN~\cite{9762321} & Binary   &   80.1     &    311  & - & -  & -    } \\
\revm{CMOS-BNN~\cite{knag2020617} & Binary   &   92.0     &    617  & - & -  & -    } \\
Ours (VGG-Small)  & Binary  &  91.7     &  $1.9\times10^5 $  & $4.8\times10^2$ & $6.2\times10^{-3}$  & 2.0      \\
Ours (VGG-Small)  & Binary  &  90.6     &  $3.8\times10^5 $  & $9.5\times10^2$ & $6.3\times10^{-3}$  & 3.9     \\
Ours (VGG-Small)   & Binary  &  89.2     & $1.5\times10^6 $   & $3.8\times10^3$ & $6.4\times10^{-3}$  & 15.2    \\
Ours (VGG-Small)   & Binary  &  87.4     &  $6.8\times10^6 $  & $1.7\times10^4$ & $7.6\times10^{-3}$  & 47.4   \\
Ours (ResNet-18)  & Binary  &  92.2     &  $1.9\times10^5 $  & $4.8\times10^2$ & $6.2\times10^{-3}$  & 2.2    \\
\bottomrule
\end{tabular}
} % scalebox
\label{tab:crossbar-array11}
% \vspace{-0.5em}
\end{table*}

\subsection{Sensitivity Analysis on Relationship between SC Bit-stream Length and Accuracy}
\label{sec:ablation_SC_length}

\begin{figure}[t]
  \centering
\includegraphics[width=0.8\linewidth]{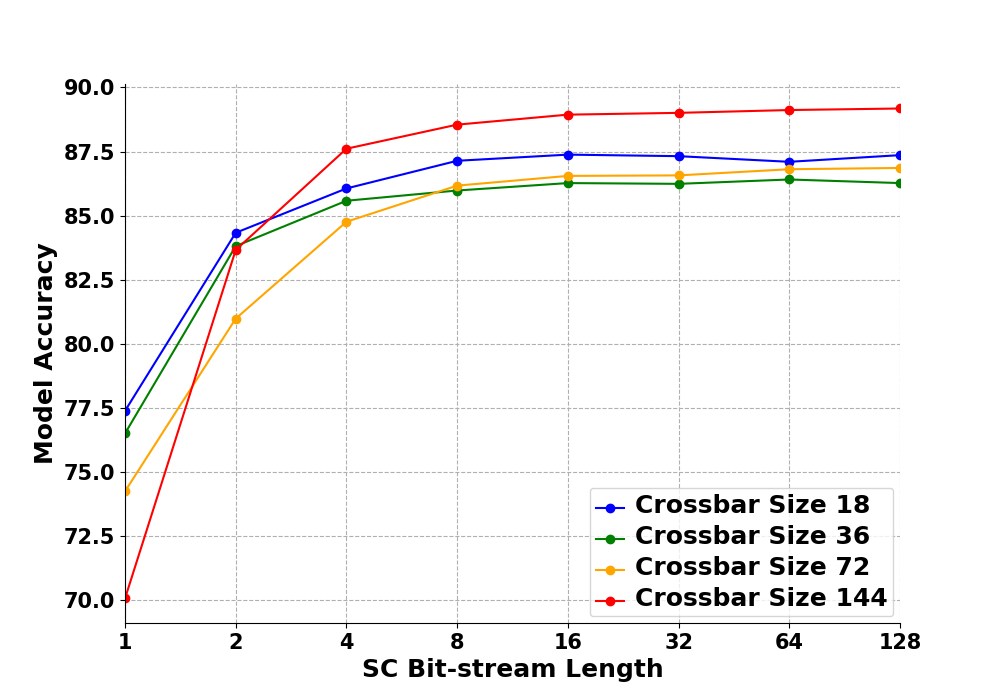}
% \vspace{-1em}
  \caption{Relationship between SC bit-stream length with model accuracy. VGG-small trained on CIFAR-10 with four different crossbar sizes are deployed in the comparison. The $\Delta I_{in}$ is set to be 2.4 $\mu A$ in this experiment.}
  \label{fig:sc_num}
  % \vspace{-1em}
\end{figure}

To choose a proper bit-stream length, we conduct a series of experiments to explore its behavior on model accuracy.
As shown in Fig~\ref{fig:sc_num}, we use VGG-small training on CIFAR-10 as an example, four different crossbar sizes are incorporated in the comparison. 
We observe that, as the SC bit-stream length increases, the model accuracy improves a lot at the beginning, but maintains a stable value after the SC bit-stream length reaches 16$\sim$32. 
Keep increasing the SC bit-stream length over 32 will not have considerable accuracy improvements but will result in a longer computing time.

\begin{figure}[h]
  \centering
  \includegraphics[width=1.0\linewidth]{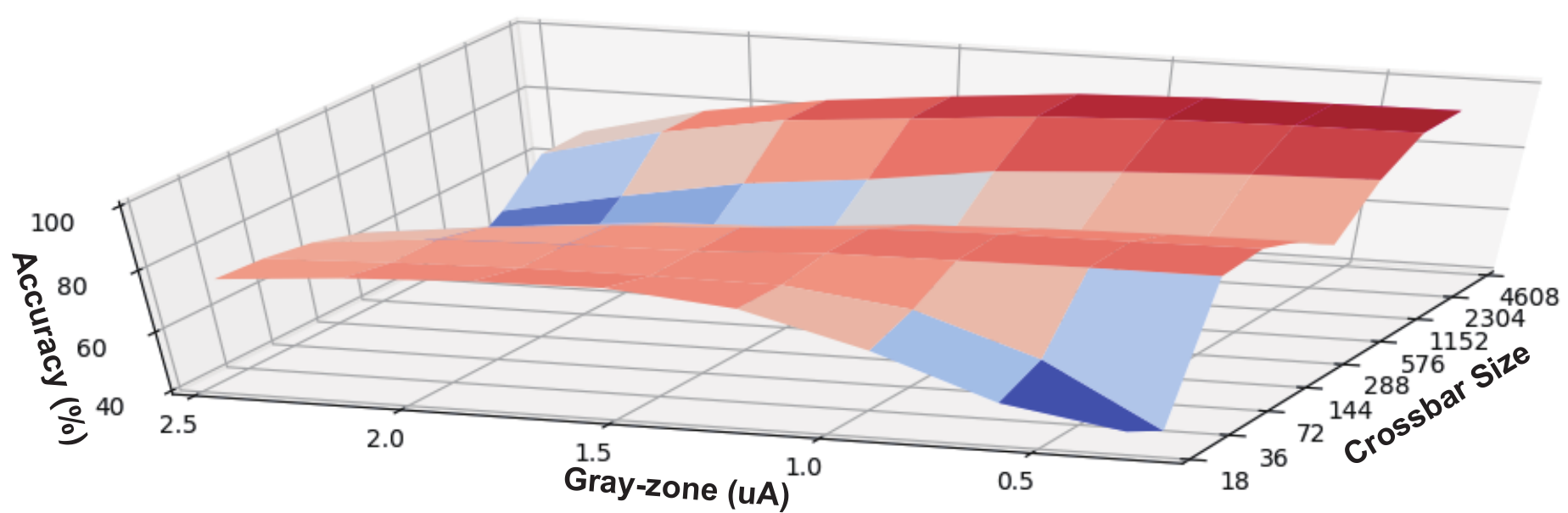}
  \caption{Accuracy distribution in two dimensions of Gray-zone $\Delta I_{in}$ and crossbar size. The stochastic bit-stream length used here is 1. }
  \label{fig:accuracy-1}
  % \vspace{-1em}
\end{figure}

\subsection{The Overall Comparison among Different $\Delta I_{in}$ Crossbar Size Configurations}

% \zg{SC number does not have to be large due to BNN feature.}
Here, we do a series of experiments to prove our methodology. VGG-small is used to be trained on CIFAR-10 dataset for these experiments. 

Using bit-stream length as 1 for the example, the overall accuracy comparison among different $\Delta I_{in}$ and crossbar size is shown in Figure~\ref{fig:accuracy-1}, where the x-axis, y-axis, and z-axis represent the values of $\Delta I_{in}$, crossbar size $C_s$ and model accuracy, respectively. As we can see, the accuracy distribution represents a close relationship to both $\Delta I_{in}$ and $C_s$. The growth trend between accuracy with one of the configurations changes largely when another one is modified. This behavior brings in multiple accuracy peaks within the whole accuracy distribution, which matches what we predicted in Section~\ref{sec:Configuration Optimization}. Using hardware benchmarks, e.g., energy consumption, efficiency, to constraint crossbar size, a comprehensive optimization can be conducted as mentioned in Section~\ref{sec:Configuration Optimization} within the target distribution area to find the local optimal solution.

% \begin{figure*}[h]
%   \centering
%   \includegraphics[width=1.0\linewidth]{Fig/accuracy-32.png}
%   \caption{ \zg{bit-stream-length+crossbar size+$\Delta I_{in}$}.}
%   \label{fig:accuracy-32}
% \end{figure*}
\subsection{Device Level Comparison with Cryogenic CMOS Technique}

\revm{In addition to superconducting devices, there has been a notable investigation into cryogenic devices based on CMOS technology, which presents itself as a viable alternative solution. These Cryo-CMOS devices offer the potential to enhance the energy efficiency of computer systems by capitalizing on diminished leakage currents and wire latency. A variety of endeavors have been undertaken in the realm of cryogenic CMOS-based research to bolster the overall performance metrics of hardware infrastructure~\cite{prasad2022cryo,saligram202164,saligram2021cryomem,nibhanupudi2021ultra,beckers2017cryogenic,alam2023cryogenic}.}

\revm{In the modern landscape of cryogenic computing, a prevalent objective encompasses the attainment of two distinct low-temperature thresholds, 77K and 4K, achieved by applying Liquid Nitrogen (LN) and Liquid Helium (LHe), respectively. Unlike superconducting computation that thrives at 4K level temperature, 77K temperature is more actively considered for cryogenic CMOS-based design to save the cooling consumption. According to~\cite{prasad2022cryo,saligram202164,saligram2021cryomem,nibhanupudi2021ultra,beckers2017cryogenic,alam2023cryogenic}, for 77 K, the cooling consumption is approximately 9.65 times the device consumption, and the 77K Cryo-CMOS can achieve about 1.5 times the energy efficiency of the conventional room temperature CMOS.}

\begin{figure}[h]
  \centering
  \includegraphics[width=1.0\linewidth]{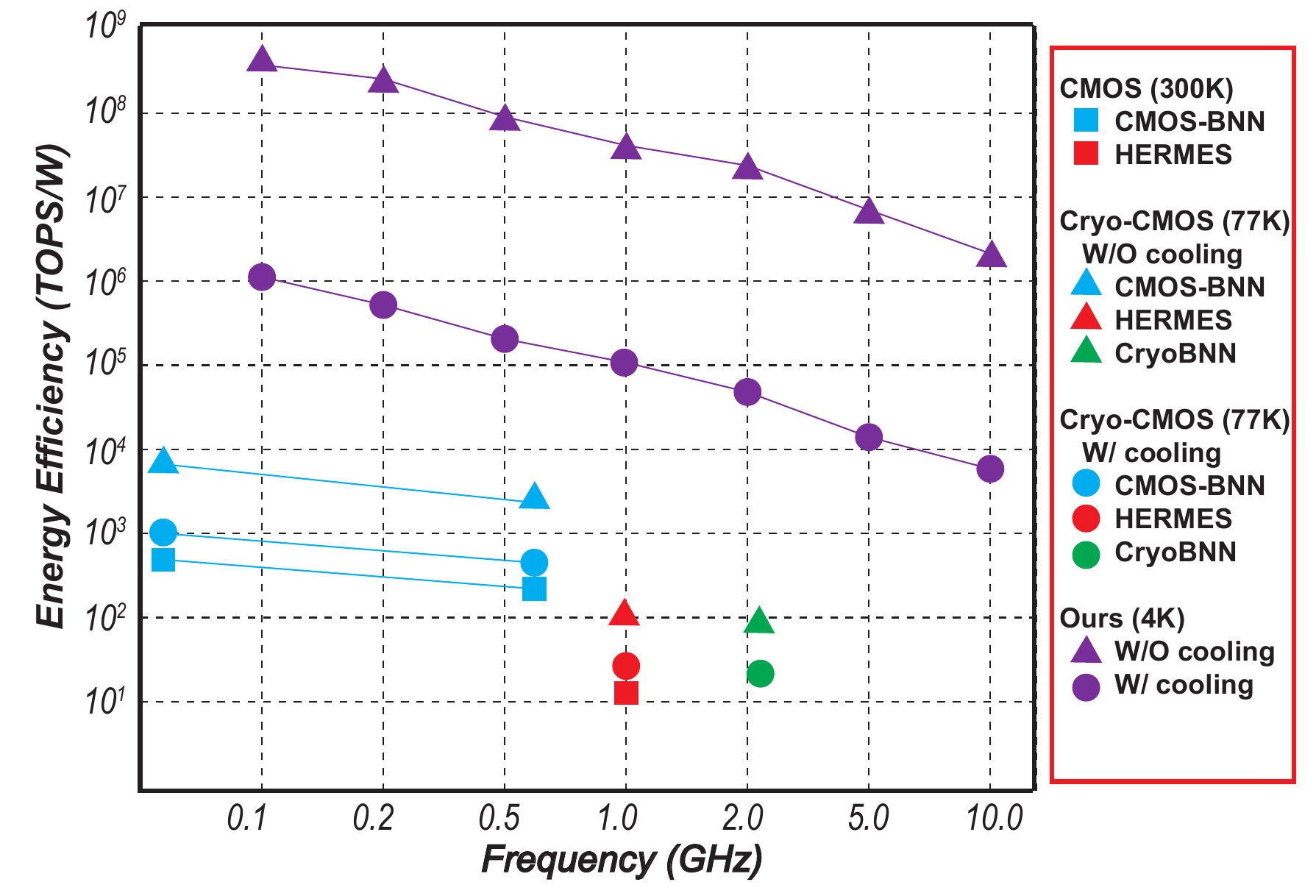}
  \caption{\revm{Comparison with room/low-temperature CMOS techniques according to energy efficiency and frequency. Among them, Cryo-CMOS counterpart results of CMOS-BNN~\cite{knag2020617} and HERMES~\cite{khaddam2021hermes} are based on estimation, Cryogenic result of CryoBNN is from~\cite{fu2022jbnn}.}}
  \label{fig:cryoCMOS}
\end{figure}

\revm{According to our device level simulation, we observe that lower frequency can generally achieve higher energy efficiency. To make a comprehensive comparison, we test our AQFP-based device under different frequencies from 0.1GHz to 10.0GHz. CMOS-BNN (1.4MHz, 622MHz)~\cite{knag2020617}, HERMES (1GHz)~\cite{khaddam2021hermes}, CryoBNN (2.24GHz)~\cite{fu2022jbnn}, and their corresponding Cryo-CMOS counterparts are incorporated in the comparison as shown in Fig,~\ref{fig:cryoCMOS}. We consider both the energy efficiency with/without cooling consumption in Cryo-CMOS and our AQFP framework. As illustrated in Fig.~\ref{fig:cryoCMOS}, in contrast to Cryo-CMOS, our approach consistently attains approximately four orders of magnitude superior energy efficiency when solely accounting for device consumption, and achieves a notable enhancement of two to three orders of magnitude in energy efficiency when factoring in cooling consumption.}

% The latest CrypCMOS shows that considering the cooling consumption, 77K CrypCMOS can achieve about 1.5 times the energy efficiency of the conventional room temperature CMOS. Since this paper doesn’t show the specific energy efficiency number, we find another new CMOS technique [A2] achieving 617 TOPs/W on BNN model. If we assume that the second one is compatible with the first cooling technique, the low-temperature energy efficiency can achieve 925.5 TOPs/W. Compared with it, we can still achieve about 50000 and 125 times better energy efficiency without/with cooling consumption, respectively. We also add this comparison to the experiment section in our revision.

\subsection{Optimization Result}

\begin{table}[h] 
\centering
% \ra{1.2} % arraystretch
\caption{Comparison with RSFQ-JBNN, ERSFQ-JBNN, CMOS-based SyncBNN, SC-AQFP, and our implementation (MLP) on MNIST Dataset.}
\label{tab:result}
% \resizebox{0.8\columnwidth}{!}{
% \begin{tabular}{@{}ccc@{}}
\scalebox{0.9}{
\begin{tabular}{c|cccc}
\toprule
\multirow{2}*{\textbf{Design}} & \multirow{2}*{\textbf{Accuracy}}   & \multicolumn{2}{c}{\textbf{Energy Efficiency (TOPS/W)} }   \\ 
& & \textbf{W/O Cooling} & \textbf{W/ Cooling} \\
\midrule
SyncBNN~\cite{fu2022jbnn} & 98.4 & 36.6 & 36.6\\
RSFQ~\cite{fu2022jbnn} & 97.9 & $2.4\times10^3 $ & 8.1\\
ERSFQ~\cite{fu2022jbnn} & 97.9  & $1.5\times10^4 $ &  50.0\\
SC-AQFP~\cite{cai2019stochastic} & 96.9  & $9.8\times10^3$ & 24.5   \\
Ours    &   98.1    &  $1.5\times10^6 $  &  $3.8\times10^3$     \\
\bottomrule
\end{tabular}}
\label{tab:mnist}
% \vspace{-0.5em}
\end{table}

As shown in Table~\ref{tab:crossbar-array11},~\M~optimizes the model accuracy according to the given energy efficiency constraints.
\revm{For CIFAR-10, we provide the result compared with DDN~\cite{chen2014dadiannao}, CMOS-BNN~\cite{knag2020617},  IMB~\cite{kim2019memory} and STT-BNN~\cite{9762321}. DDN is a representative CMOS-based digital accelerator. CMOS-BNN is a BNN accelerator based on 10-nm FinFET CMOS under low frequency, 13MHz (thus has a higher energy efficiency). IMB uses resistive memory crossbar Array (RCA) with RRAM architecture to implement BNN computation. STT-BNN combines spin-transfer torque magnetoresistive Random Access Memory (MRAM) with BNN to improve energy efficiency. Besides these works, recent work~\cite{byun2020cryocore} explore the low temperature CMOS (77K), which may potentially achieve 1.5 times overall better energy efficiency compared with the conventional room temperature CMOS.}
% IMB which can compute a vector-matrix multiplication in a single time step. If an input vector is mapped to input voltages on the memory rows and the weight matrix is mapped to conductance values of each device, currents flowing through the columns are the results of a vector-matrix multiplication. The frequency of IMB is 100MHz.
\rev{For full-precision VGG-small model, DDN can achieve 92.5$\%$ top-1 accuracy with 0.45 TOPS/W energy efficiency. Our~\M~achieves $4.2\times10^5$ times better energy efficiency with a similar level of accuracy (92.2\%) on the same model.}
\rev{Compared with IMB with BNN model,~\M~has a much higher frequency 5GHz, $7.8\times10^4$ times of higher energy efficiency with similar model accuracy. When we loosen the efficiency constraint,~\M ~can achieve 91.7\% and 92.2\% top-1 model accuracy on VGG-small and ResNet-18, respectively, with the energy efficiency of $1.9\times10^5$ TOPS/W. }
% \rev{For ImageNet dataset, DDN achieves 0.45 TOPS/W energy efficiency on full-precision Mobilenet-V2-0.35 model, which has 4.6$\times$ larger model size compared with our binary ResNet-18. With the same level of model accuracy, CBNN~\cite{andri2021chewbaccann}, a CMOS-based BNN accelerator, achieves 55$\times$ better energy efficiency compared with DDN. While our approach obtains 59.1\% and 60.3\% top-1 accuracy with $1.9\times10^5$ TOPS/W energy efficiency on VGG-16 and ResNet-18, respectively.}
\rev{The cooling cost for typical superconducting digital circuits is about 400$\times$ the chip power dissipation~\cite{holmes2013energy}. Even considering cryo energy,~\M~still shows 205.8$\times$ higher energy efficiency compared to IMB under the same level of accuracy.
}

\rev{To compare with JBNN~\cite{fu2022jbnn} and SC-AQFP~\cite{cai2019stochastic}, we test our approach on MNIST dataset. As shown in Table~\ref{tab:mnist}, SyncBNN, RSFQ, and ERSFQ are CMOS-based, RSFQ-based, and ERSFQ-based BNN accelerators in JBNN paper~\cite{fu2022jbnn}. SC-AQFP is the AQFP-based pure stochastic computing accelerator. We use the same model architecture (MLP) as shown in JBNN~\cite{fu2022jbnn}. With the same BNN model, whether considering the cooling energy or not, our approach consistently achieves two to four orders of magnitude better energy efficiency compared with CMOS-based, RSQ-based, and ERSFQ-based accelerators with similar accuracy. Compared with SC-AQFP, which processes pure stochastic computing on AQFP devices, our approach achieves 153$\times$ better energy efficiency for both cooling/non-cooling situations with $2\%$ better top-1 accuracy.
}

\section{Discussion}\label{sec:disc}
\revm{In addition to AI-focused accelerators, the proposed AQFP technology can also be employed for conventional general-purpose computing to cater to a variety of application scenarios. The AQFP technology boasts a standard cell library designed for different manufacturing processes, such as Japan AIST 4-layer process HSTP ~\cite{hstp} and US MIT-LL 8-layer Nb process SFQ5ee ~\cite{mitprocess}, presenting a rich assortment of over 80 cells \cite{coldflux}. This includes 3- and 5-input logic gates, signal-driving boosters, and refined interfaces across various superconducting logic families. }

\revm{The development of a comprehensive EDA toolchain—from logic synthesis to placement and routing—is specifically tailored for this standard cell library. Digital modelling and a synthesis flow for cell-based AQFP structural circuit generation were proposed in 2017, which can be seen as the earliest attempt towards AQFP design automation~\cite{tas2017}. This synthesis flow is further tailored to support more AQFP features by different research groups in~\cite{huangiccad2021,epfl_aspdac,fuaspdac2023}. For the development of placement and routing, T. Tanaka et al. proposed a framework using a genetic algorithm (GA) for placement and a left-edge channel routing scheme in 2019~\cite{tas2019}, whereas Y. Chang et al. proposed another framework adapting a learning-based placer to minimize the runtime overhead in 2020~\cite{tinghua_iccad}. H. Li et al. have developed a different tool using a negotiation-based A* router, targeting processes allowing multiple routable metal layers~\cite{date2021}. System-level performance analysis has been successfully conducted using the aforementioned EDA framework \cite{chen2019adiabatic}. These results ensure that AQFP technology is compatible with conventional logic and memory design, including but not limited to microprocessor, register file and random-access memory. }

\revm{Moreover, by amplifying superconducting signals to voltage levels, specially designed on-chip interfaces between AQFP and conventional CMOS technologies have been implemented and demonstrated ~\cite{jld}. This paves the way for the system to be employed in broader applications, including supercomputing, cloud computing, and secure computing.}
\section{Conclusion}\label{sec:concl}

In this paper, \rev{we first make the analysis of the randomized behavior of AQFP buffer and current attenuation feature of AQFP, then propose the randomized-aware BNN training algorithm  effectively integrating the randomized behavior into the BNN training process.} 
\rev{To solve the intermediate results accumulation problem as well as preserve the model accuracy, we inspiringly convert the randomized output of the neuron circuit to the stochastic computing domain and propose a novel stochastic computing-based accumulation module.}
Finally, we propose an algorithm-hardware co-optimization method and batch normalization matching to close the gap between software with hardware. The clocking scheme adjustment-based circuit optimization is also applied to improve the overall performance.
Based on our algorithm-hardware co-optimization the hardware configurations of our AQFP-based randomized BNN accelerator, including crossbar synapse array size, stochastic computing bit-stream length, and ``gray-zone'' width of AQFP buffer by comprehensively considering power consumption, energy efficiency, and hardware computing error, are jointly optimized.

%%%%%%% -- PAPER CONTENT ENDS -- %%%%%%%%
\section*{ACKNOWLEDGMENTS}
This study was supported by JST PRESTO Program (Grant No. JPMJPR19M7), FOREST Program (Grant Number JPMJFR226W, Japan), and the NSF Expedition program CCF-2124453, NSF CCF-2008514.

%%%%%%%%% -- BIB STYLE AND FILE -- %%%%%%%%
\bibliographystyle{IEEEtranS}
\bibliography{refs}
%%%%%%%%%%%%%%%%%%%%%%%%%%%%%%%%%%%%

\end{document}